\documentclass[aps,prd,nofootinbib,showpacs,preprintnumbers,amsmath,amssymb,superscriptaddress,groupedaddress,floatfix]{revtex4-2}
\usepackage[pdftex,colorlinks=true,bookmarksopen=true]{hyperref}
\usepackage[utf8]{inputenc}
\usepackage{amsmath,amssymb}
\usepackage{physics}
\usepackage{slashed}
\usepackage{color}
\usepackage{graphicx}
\usepackage{tikz}
\usetikzlibrary{decorations.markings}
\usepackage{bbm}

\usepackage{xcolor}
\definecolor{tabblue}{HTML}{1F77B4}
\definecolor{tabgreen}{HTML}{2CA02C}
\definecolor{taborange}{HTML}{FF7F0E}
\definecolor{tabred}{HTML}{D62728}
\definecolor{dark1}{HTML}{D95F02}
\definecolor{dark2}{HTML}{7570B3}

\renewcommand{\tr}[1]{\ensuremath{\text{tr}\left[#1\right]}}
\newcommand{\trc}[1]{\ensuremath{\text{tr}_c\left[#1\right]}}
\newcommand{\trs}[1]{\ensuremath{\text{tr}_s\left[#1\right]}}

\begin{document}
\title{
    Spectral analysis for nucleon-pion and nucleon-pion-pion states in both parity sectors using distillation with domain-wall fermions
}
\newcommand{\regensburg}{Fakutät für Physik, Universität Regensburg, Universitätsstraße 31, 93040 Regensburg}

\author{A.~\surname{Hackl}}\email{andreas.hackl@ur.de}\affiliation{\regensburg}
\author{C.~Lehner}\affiliation{\regensburg}
\thanks{Present address: \rege}
\date{\today}

\begin{abstract}
    We present a study using the distillation method to analyze the spectra of nucleon, nucleon-pion, and nucleon-pion-pion states in the positive-parity sector, as well as nucleon and nucleon-pion states in the negative-parity sector. The study uses seven domain-wall fermion ensembles with varying pion masses ($m_\pi = 139 - 279~\text{MeV}$), lattice spacings ($a^{-1} = 1.730~\text{GeV}$ and $a^{-1}=2.359~\text{GeV}$) and volumes ($m_\pi L = 3.8 - 7.5$). To address the large number of contractions in this project, we implemented an algorithm to automate the contraction of nucleon-pion correlation functions that contain an arbitrary number of pions. In the positive parity sector, we extrapolate the nucleon mass to the physical point. This study demonstrates the effectiveness of the distillation method for baryonic quantities with a focus on multi-hadronic states and establishes a foundation for future work. 
\end{abstract}

\maketitle

\section{Introduction}

Neutrino experiments of the next generation, such as LBNF/DUNE \cite{DUNE:2015lol} and HyperK \cite{Hyper-Kamiokande:2018ofw}, require a precise prediction of the nucleon-neutrino cross section in the energy range from few MeV to GeV \cite{Formaggio:2012cpf}. This energy region overlaps with the resonance region in which incoming (anti)neutrinos excite the target nucleon to resonance states such as the $\Delta(1232)$ and the roper resonance $N(1440)$ which decay to $N\pi$ and $N\pi\pi$ states.  In general, neutrino experiments do not use single nucleons as targets but large nuclei in which the individual nucleons are embedded. So far, a direct lattice QCD study for large nuclei is not feasible; however, they can be treated using effective field theory, for which lattice QCD can provide non-perturbative first principle calculations of nucleon inputs \cite{Kronfeld:2019nfb, Meyer:2022mix, Ruso:2022qes}. One of the most prominent nucleon inputs is the nucleon axial-vector form factor and its zero-momentum limit, the nucleon axial-vector charge. 

An important challenge for the nucleon axial-vector form factor is a substantial systematic error from $N\pi$ states \cite{B_r_2015, B_r_2018} as predicted by chiral perturbation theory. Multiple publications address this issue by using different approaches \cite{Bali:2018qus, RQCD:2019jai, Jang:2019vkm, Park:2021ypf, Djukanovic:2022wru, He:2021yvm}. Most studies conducted so far use three-quark interpolating operators. However, there are first approaches to also include 5-quark operators \cite{Barca_2023, Alexandrou:2023ajp, Alexandrou:2024tin} and 7-quark operators \cite{Grebe:2023tfx}, which are designed to be multi-hadronic operators with an increased overlap with $N\pi$ and $N\pi\pi$ states. Combining multi-hadronic and standard nucleon operators has been shown to be effective in creating nucleon operators without $N\pi$ overlap \cite{Barca_2023, Alexandrou:2024tin} and also offers a way to construct excited state operators without ground state contributions. Such excited state operators can be vital for computations of matrix elements containing excited nucleon states and resonances which are relevant for processes in the resonant regime of neutrino-nucleon interactions \cite{Simons:2022ltq}.

This study aims to contribute to this effort by providing a generalized eigenvalue problem analysis of nucleon, $N\pi$, and, for the first time, $N\pi\pi$ states in the positive parity channel as well as an analysis in the negative sector using $N^-$ and two different $N\pi$ operators.

Our analysis uses the distillation method \cite{Peardon_2009} to compute the multi-hadronic correlation functions effectively. Due to its high-mode suppression, distillation offers a smearing procedure that is fitted for creating operators with strong overlap with the low-mode states of the theory. A disadvantage of distillation is its substantial computational cost of computing the corresponding distillation objects such as the perambulators. However, once these objects are computed and stored, it is affordable to compute general $n$-point functions with these objects. This work was only possible because we have access to a large database of distillation data generated by the RBC/UKQCD collaborations in their effort to measure the hadronic vacuum polarization of $g-2$ \cite{Bruno:2019nzm,Bruno:2023pde,RBC:2024fic} which, however, means that the distillation setup was not optimized for baryon physics.

Throughout this study, we have to compute correlation functions involving multiple particles. Some of those exhibit many contractions, which require automatic execution. We propose a way to automate this process of contracting correlation function and even generate code to perform the required tensor projections to calculate correlation functions using the distillation method \cite{Hackl_AutoWick_2024}.

This publication should be seen as an exploratory work to test the effectiveness of the distillation method for baryonic quantities for which the nucleon spectrum is a good starting point. We want to emphasize that there is previous work using distillation of baryonic quantities \cite{Lang_2013, Lang_2017, Egerer_2019}. However, all previous studies use either large pion masses, small physical volumes, or fermion discretizations without good chiral properties.

We apply All Mode Averaging \cite{DeGrand_2005, Bali_2010, Blum_2013, Shintani_2015} to reduce the computational cost. A further reduction in computation costs is achieved by using the Wigner-Eckart theorem to relate different correlation functions with each other and computing only independent correlation functions \cite{Barca_2023}.

The paper is structured as follows: In the next section \autoref{sec:methods}, we describe the methods used in this study, including the operator basis, the generalized eigenvalue problem analysis, and the distillation method. In \autoref{sec:computational}, we present the computational details consisting of a short description of the RBC/UKQCD collaboration ensembles and the AMA method. Furthermore, the section also includes an explanation of the algorithm to automatically contract the correlation function used in this study. In \autoref{sec:positive-parity}, we present the spectral analysis results for the nucleon, $N\pi$, and $N\pi\pi$ states in the positive parity sector. In this sector, we also project to the physical world by identifying the finite volume corrections and project to the physical pion and kaon masses and the continuum. In \autoref{sec:negative-parity}, we perform the spectral analysis of the nucleon and $N\pi$ states in the negative parity sector. We conclude with a discussion in \autoref{sec:conclusion}.

\section{Methods}\label{sec:methods}
This section explains the methods we used throughout our analysis, starting with our operator choice in the positive and negative parity sectors. We further describe the generalized eigenvalue problem (GEVP) analysis, which obtains the nucleon spectrum for the individual ensembles. At the end of this section, we explain how we use the distillation method to construct a matrix of correlation functions used in the GEVP analysis.
\subsection{Interpolating operator}
The set of interpolating operators for the positive parity channel in this project is the nucleon $N$ and isospin projected multi-hadronic p-wave $N\pi$ and s-wave $ N\pi\pi$ operators. For the negative parity channel, we replace the s-wave $N\pi\pi$ operator with an s-wave $N\pi$ operator, which naturally has negative parity. This paper will only consider $I=\frac12$, $I_z=\frac12$ isospin states without loss of generality. For the single nucleon operators, we are using the standard interpolating operator 
\begin{align}\label{eq:nuc_op}
    \mathcal{O}_N(\boldsymbol p, t)_\alpha &= \sum_{\boldsymbol{x}} e^{i \boldsymbol{p} \cdot \boldsymbol{x}} \varepsilon^{abc} \psi^a_\alpha(x) \left(u^b(x)^T C \gamma^5 d^c(x)\right),
\end{align}
where $x$ denotes the combination of spatial and time position $x = (\boldsymbol x, t)$ and $\psi$ is chosen to be $u$ for protons and $d$ for neutrons. The parity of the operator is fixed by multiplying it with the corresponding parity projection $P^\pm = \frac12 \left(\mathbbm{1} + \gamma^t\right)$ to obtain \begin{equation}
    \mathcal{O}_N^\pm(\boldsymbol{p}, t)_\alpha = P^\pm_{\alpha\beta} \mathcal{O}_N(\boldsymbol{p}, t)_\beta.
\end{equation} 
For $\boldsymbol p = \boldsymbol{0}$, both operators transform in the $G_{1,u/g}$ irreducible representation of the double cover of the octahedral group $O_h^d$. 
For the pion interpolating operators, we use
\begin{align}
    \mathcal{O}_{\pi^+}(\boldsymbol{p}, t) &= \sum_{\boldsymbol{x}} e^{i \boldsymbol{p} \cdot \boldsymbol{x}} \bar{d}(x) \gamma^5 u(x) \\
    \mathcal{O}_{\pi^0}(\boldsymbol{p}, t) &= \sum_{\boldsymbol{x}} e^{i \boldsymbol{p} \cdot \boldsymbol{x}} \frac{\bar{u}(x) \gamma^5 u(x) - \bar{d}(x) \gamma^5 d(x)}{\sqrt2}\\
    \mathcal{O}_{\pi^-}(\boldsymbol{p}, t) &= \sum_{\boldsymbol{x}} e^{i \boldsymbol{p} \cdot \boldsymbol{x}} \bar{u}(x) \gamma^5 d(x).
\end{align} 
In the rest frame, the pion operators transform in the irreducible representation $A_{1,u}$ of $O_h^d$. \\
The relevant multi-hadronic operators are then linear combinations of products of nucleon and pion interpolating operators. In this project, we use three different operator products. The first one is the nucleon-pion operator in s-wave, i.e., both particles are at rest \cite{Lang_2013}, 
\begin{equation}\label{eq:npi_swave}
    \mathcal{O}^{(000)}_{N\pi,G_{1u}}(\boldsymbol{0}, t)_\alpha = \gamma^5_{\alpha \beta} \mathcal{O}^+_N(\boldsymbol{0}, t)_\beta \mathcal{O}_\pi(\boldsymbol{0}, t),
\end{equation}
where $(000)$ is introduced to label that both particles are at rest. $\mathcal{O}_N$ and $\mathcal{O}_\pi$ denote general nucleon and pion operators, respectively. We further need to project the nucleon-pion operator to the isospin $(I, I_z) = (\frac12,\frac12)$ state, which is done by the linear combination 
\begin{equation}\label{eq:isospin_np}
    \mathcal{O}^{(000)}_{N\pi,G_{1u},\frac12 \frac12} = \sqrt{\frac{2}{3}} \mathcal{O}^{(000)}_{n\pi^+,G_{1u}} + \sqrt{\frac{1}{3}} \mathcal{O}^{(000)}_{p\pi^0,G_{1u}}.
\end{equation}
Since the overall parity of this state is negative, this multi-hadronic state only contributes to the negative parity channel. Giving both operators a back-to-back momentum yields the possibility of projecting to a positive parity operator. In the following, we consider the smallest possible momenta on the lattice, $\hat{p}_i = \frac{2\pi}{L_s} \hat{e}_i$, where $L_s$ denotes the number of spatial lattice sites and $\hat{e}_i$ is the unit vector in $i$-th direction. For these momenta, the projection on $G_{1g}$ is 

\begin{subequations}
\begin{align}\label{eq:npi_pos_parity}
\begin{aligned}
    \mathcal{O}^P_{N\pi, G_{1g}}(\boldsymbol 0, t)_\uparrow &= \mathcal{O}^+_N(\hat{p}_x, t)_\downarrow \mathcal{O}_\pi(-\hat{p}_x, t) - \mathcal{O}^+_N(-\hat{p}_x, t)_\downarrow \mathcal{O}_\pi(\hat{p}_x, t) \\
    &-i \mathcal{O}^+_N(\hat{p}_y, t)_\downarrow \mathcal{O}_\pi(-\hat{p}_y, t) + i \mathcal{O}^+_N(-\hat{p}_y, t)_\downarrow \mathcal{O}_\pi(\hat{p}_y, t)\\
    &+\mathcal{O}^+_N(\hat{p}_z, t)_\uparrow \mathcal{O}_\pi(-\hat{p}_z, t) - \mathcal{O}^+_N(-\hat{p}_z, t)_\uparrow \mathcal{O}_\pi(\hat{p}_z, t), \\
\end{aligned} \\
\begin{aligned}
    \mathcal{O}^P_{N\pi, G_{1g}}(\boldsymbol 0, t)_\downarrow &= \mathcal{O}^+_N(\hat{p}_x, t)_\uparrow \mathcal{O}_\pi(-\hat{p}_x, t) - \mathcal{O}^+_N(-\hat{p}_x, t)_\uparrow \mathcal{O}_\pi(\hat{p}_x, t) \\
    &+ i \mathcal{O}^+_N(\hat{p}_y, t)_\uparrow \mathcal{O}_\pi(-\hat{p}_y, t) - i \mathcal{O}^+_N(-\hat{p}_y, t)_\uparrow \mathcal{O}_\pi(\hat{p}_y, t)\\
    &-\mathcal{O}^+_N(\hat{p}_z, t)_\downarrow \mathcal{O}_\pi(-\hat{p}_z, t) + \mathcal{O}^+_N(-\hat{p}_z, t)_\downarrow \mathcal{O}_\pi(\hat{p}_z, t), \\
\end{aligned}
\end{align}
\end{subequations}
where the $\uparrow$ and $\downarrow$ symbols represent even and odd spins, respectively \cite{Prelovsek:2016iyo, Lang_2017, Barca_2023}. The back-to-back momenta create a vector-like structure, transforming on the lattice in the irreducible representation $T_1$. Since the $SU(2)$-irrep $J=1$ subduces among others to $T_1$, we call such states p-wave states. This momentum configuration also allows for a projection onto $G_{1u}$, contributing to the negative parity channel. Following \cite{Prelovsek:2016iyo} we get 
\begin{equation}\label{eq:npi_neg_parity}
    \begin{aligned}
        \mathcal{O}^{(001)}_{N\pi, G_{1u}}(\boldsymbol 0, t)_{\uparrow/\downarrow} &= \mathcal{O}_N^+(\hat{p}_x, t)_{\uparrow/\downarrow} \mathcal{O}_\pi(-\hat{p}_x, t) + \mathcal{O}_N^+(-\hat{p}_x, t)_{\uparrow/\downarrow} \mathcal{O}_\pi(\hat{p}_x, t) \\
        &+ \mathcal{O}_N^+(\hat{p}_y, t)_{\uparrow/\downarrow} \mathcal{O}_\pi(-\hat{p}_y, t) + \mathcal{O}_N^+(-\hat{p}_y, t)_{\uparrow/\downarrow} \mathcal{O}_\pi(\hat{p}_y, t) \\
        & + \mathcal{O}_N^+(\hat{p}_z, t)_{\uparrow/\downarrow} \mathcal{O}_\pi(-\hat{p}_z, t) + \mathcal{O}_N^+(-\hat{p}_z, t)_{\uparrow/\downarrow} \mathcal{O}_\pi(\hat{p}_z, t),
    \end{aligned}
\end{equation}
where $(001)$ labels the back-to-back momentum configuration with momenta $\hat{p}_i = \frac{2\pi}{aL} \hat{e}_i$. A short group theoretical discussion about the derivation of both back-to-back operators is given in App.~\ref{sec:interpolating_operator_construction}.

The next multi-hadronic operator we are considering is the s-wave $N\pi\pi$ operator, defined by
\begin{equation}\label{eq:npp_swave}
    \mathcal{O}^S_{N\pi\pi,G_{1g}}(\boldsymbol{0}, t)_\mu = \mathcal{O}_N(\boldsymbol{0}, t)_\mu \mathcal{O}_\pi(\boldsymbol{0}, t) \mathcal{O}_\pi(\boldsymbol{0}, t).
\end{equation}
The isospin projection for the $N\pi\pi$ state without momenta is given by 
\begin{equation}
    \mathcal{O}^S_{N\pi\pi,G_{1g},\frac12 \frac12} = \frac{2}{\sqrt 5} \mathcal{O}^S_{p\pi^+ \pi^-, G_{1g}} - \frac{1}{\sqrt{5}} \mathcal{O}^S_{p\pi^0 \pi^0, G_{1g}}.
\end{equation}
To summarize, we use for the GEVP analysis in negative parity channel the operator basis $\mathcal{O}_p^-, \mathcal{O}_{N\pi,G_{1u}}^{(000)}, \mathcal{O}_{N\pi,G_{1u}}^{(001)}$ and $\mathcal{O}_p^+, \mathcal{O}_{N\pi,G_{1g}}^P, \mathcal{O}_{N\pi\pi, G_{1g}}^S$ for the positive parity channel. 
\subsection{Generalized Eigenvalue Problem Analysis}
In our spectral analysis, we use the Generalized Eigenvalue Problem (GEVP) analysis \cite{alpha_2009}, where we consider a matrix of correlation functions with  entries
\begin{equation}
    C_{ij}(t) = \left\langle\mathcal{O}_i(t) \mathcal{O}^\dagger_j(0) \right\rangle, \quad \text{for } i,j = 1, \dots, N
\end{equation}
using the projected interpolating operators from the previous section. In Euclidian space-time, the correlation functions can be expressed as a sum over decaying exponentials representing the tower of states with the same quantum numbers as the interpolating operators and ascending energies $E_n$, yielding the equation 
\begin{equation}
    C_{ij}(t) = \sum_n \mel{0}{\hat{\mathcal{O}}_i}{n}\mel{n}{\hat{\mathcal{O}}_j^\dagger}{0} e^{-E_n t}.
\end{equation}
The matrix elements $\mel{0}{\hat{\mathcal{O}}_i}{n}$ denote the overlap of the interpolating operator with the physical finite-volume state $\ket{n}$.   By solving the GEVP, i.e., 
\begin{equation}
    C(t) V_n(t,t_0) = \lambda_n(t, t_0) C(t_0) V_n(t, t_0),
\end{equation}
we can obtain the first $N$ eigenenergies via the eigenvalues $\lambda_n(t,t_0)$. The eigenvalues have the form \cite{LUSCHER1990222,alpha_2009}
\begin{equation}
    \lambda_n(t, t_0) = e^{- E_n (t - t_0)},
\end{equation}
assuming the tower of states consists only of $N$ states. The eigenenergies of the states are then computed using
\begin{equation}
    a m_{\text{eff}, n}(t) = - \log\left(\frac{\lambda_n (t+a, t_0)}{\lambda_n(t, t_0)}\right) = a E_n.
\end{equation}
In general more than $N$ states contribute to the correlation function and the excited state contributions can be approximated by
\begin{equation}
    a m_{\text{eff}, n}(t) = a E_n + \alpha_n e^{- (E_{N+1} - E_n) t} .
\end{equation}
Here we assume that $t_0 \geq t/2$ and we neglect higher excited states \cite{alpha_2009}. Our analysis chose the distance between $t$ and $t_0$ as the lattice spacing, i.e., $t - t_0 = a$. The condition $t_0 \geq t/2$ is then fulfilled for $t \geq 2a$. For $t$ where the excited state contributions are negligible, the eigenvectors $v_n(t, t_0)$ also converge to a constant value independent of $t$ and $t_0$. In the asymptotic limit, i.e., for large $t$, we can define new interpolating operators in the eigenbasis of GEVP, i.e., 
\begin{equation}
    \tilde{\mathcal{O}}_n = \sum_m V_{nm}(t, t - \Delta) \mathcal{O}_m, \quad \text{ for } t \gg a,
\end{equation}
which have vanishing overlap with all eigenstates $m \neq n$ and $m \leq N$. This new basis of operators can be used to compute other quantities, building on the previous operators, with better control over the excited state systematics. Prominent examples are the computation of form-factors \cite{Barca_2023} and scattering phase shifts \cite{Blum_2023}. However, in this work, we only focus on the GEVP analysis itself, and other quantities like the nucleon axial-vector form factor will be part of future work. 

\subsection{Distillation}
The correlation functions of this study have been constructed using the distillation method \cite{Peardon_2009}. This method uses the eigenvectors of a smearing operator, typically the Laplace operator, to construct smeared quark fields that strongly overlap with the theory's low-mode states. The eigenvectors are solutions of the eigenvalue equation
\begin{equation}
    \sum_{\boldsymbol{y}} L(\boldsymbol x, \boldsymbol{y}) V^n(\boldsymbol{y}) = V^n(\boldsymbol{x}) \lambda^n,
\end{equation}
with ascending eigenvalues, i.e., $\lambda^m \geq \lambda^n$ for $m > n$, of the three-dimensional discrete spatial Laplace operator
\begin{align}\label{eq:laplace}
    L(\boldsymbol{x}, \boldsymbol{y}) = - \delta_{\boldsymbol{x},\boldsymbol{y}} + \frac1{6a^2} \sum_i \left(U_i(\boldsymbol{x}) \delta_{\boldsymbol{x}, \boldsymbol{y}-a \boldsymbol{\hat{i}}} + U^\dagger_i(\boldsymbol{x-a\boldsymbol{\hat i}}) \delta_{\boldsymbol{x}, \boldsymbol{y}+a\boldsymbol{\hat i}}\right).
\end{align}
In our distillation setup we use a particular gauge-smearing technique described in more detail in Ref.~\cite{Bruno:2023pde}.
These eigenvectors are then used as sources for solving the Dirac equation for the quark propagator G,
\begin{equation}
    \sum_{y} \slashed{D}(x, y) G^n(t_y, t_x, \boldsymbol{y}) = V^n(x)
\end{equation}
with suppressed spin and color indices. Furthermore, we collect space and time positions in the non-bold coordinates, i.e., $x = (\boldsymbol x, t)$. Multiplying the quark propagator with the hermitian-conjugate eigenvectors at the sink yields the so-called perambulator 
\begin{equation}
    \mathcal{G}^{mn}(t_y, t_x) = \sum_{\boldsymbol{y}} V^m(t_y, \boldsymbol{y})^\dagger G^n(t_y, t_x, \boldsymbol{y}),
\end{equation}
which is of the size $N_d^2 \times 4^2$, where $N_d$ is the number eigenmodes of the Laplace operator. The perambulator has only spin indices since the color indices are contracted with each of the color indices of the eigenvectors. We can view these perambulators as propagators in the space of eigenmodes of the Laplace operators. This basis transformation comes with the cost of truncating high modes of the Laplace operator. In our analysis, we observe that we might have signs of this truncation in the negative parity sector. A useful measure to determine the width of the smearing due to the distillation approach is the profile of the distillation operator \cite{Peardon_2009},
\begin{equation}
    \Psi(\boldsymbol{r}) = \sum_{\boldsymbol{x}, t} \sqrt{\trc{\Box(\boldsymbol{x}, \boldsymbol{x} + \boldsymbol{r}, t) \Box(\boldsymbol{x} + \boldsymbol{r}, \boldsymbol{x}, t)}},
\end{equation}
where $\Box$ denotes the distillation operator $\Box(\boldsymbol x, \boldsymbol y, t)_{cc'} = \sum_{k=1}^{N_d} V^k_{c}(\boldsymbol x, t) \left[V^k_{c'}(\boldsymbol y, t)\right]^\dagger$ and $\trc{A}$ is the trace over all color indices of the operator $A$. The profiles of all ensembles used in this study are shown in \autoref{fig:profile}. The distillation operator applied to point-like quarks yields the smeared quarks 
\begin{equation}\label{eq:smeared_quark}
    q^s_{\alpha, c}(x) = \sum_{\boldsymbol{y}} \Box(\boldsymbol{x}, \boldsymbol{y}, t)_{cc'} q_{\alpha, c'}(y).
\end{equation}
For the new smeared quarks, the propagator is related to the perambulator by the basis transformation
\begin{equation}
    G^s_{\alpha \alpha', c c'}(y \vert x) \equiv \left\langle q_{\alpha, c}^s(\boldsymbol{y},t_y) \bar{q}_{\alpha', c'}^s(\boldsymbol{x}, t_x) \right\rangle = \sum_{n,m} V_c^n(\boldsymbol{y}, t_y) \mathcal{G}^{nm}_{\alpha \alpha'}(t_y, t_x) \left[V_{c'}^m(\boldsymbol{x}, t_x)\right]^\dagger, 
\end{equation}
which is obtained by inserting \autoref{eq:smeared_quark} into the definition of the smeared propagator. For these smeared propagators, the two-point pion correlation function can be expressed as
\begin{align}
    C_{\pi\pi}(t, \boldsymbol{p}) &= \left\langle \mathcal{O}_\pi(\boldsymbol{p}, t) \mathcal{O}^\dagger_\pi(\boldsymbol{p}, 0) \right\rangle \\
    &= - \left\langle \tr{\mathcal{G}^{mm'}(t,0) \mathcal{P}^{m'n'}(0, -\boldsymbol{p}) \gamma^5 \mathcal{G}^{n'n}(0, t) \gamma^5 \mathcal{P}^{nm}(t,\boldsymbol{p})} \right\rangle,
\end{align}
where $\mathcal{P}(t, \boldsymbol{p})$ denotes a momentum insertion defined by
\begin{equation}
    \mathcal{P}^{nm}_{cc'}(t, \boldsymbol{p}) = \sum_{\boldsymbol{x}} \left[V^{n}_c(\boldsymbol x, t)\right]^\dagger e^{i \boldsymbol{x} \cdot \boldsymbol{y}} V^{m}_{c'}(\boldsymbol{x}, t).
\end{equation}
The convenient property of the distillation method is that all $n$-point functions, including only mesons, can be expressed as a tensor contraction over multiple perambulators and momentum insertions. Typically, the number of Laplace modes is much smaller than the spatial volume. Hence, both can be stored on disk for relevant momenta and time positions. Moreover, contracting the perambulators and momentum insertions also requires far less compute time, enabling the computation of more complex $n$-point functions. 
The inclusion of baryons requires an additional building block in the distillation toolbox, which emerges from the proton two-point function in the contracted form,
\begin{align}
    C_{pp}(t, \boldsymbol{p}) &= \left\langle \mathcal{O}_p(\boldsymbol{p}, t) \mathcal{O}^\dagger_p(\boldsymbol{p}, 0) \right\rangle \\
    \label{eq:2pt_proton}
    &=  \sum_{\boldsymbol{x},\boldsymbol{y}} e^{i \boldsymbol{p} \cdot (\boldsymbol{y} - \boldsymbol{x})} \left\langle\tr{P^+ \trc{\left\{\trs{\mathcal{Q}^{pp}(x\vert y)} + \mathcal{Q}^{pp}(x\vert y)\right\}U(x \vert y)}} \right\rangle,
\end{align}
with the quark contraction structure
\begin{align}
    \mathcal{Q}^{pp}_{\alpha\beta,a' a}(x\vert y) &= \mathcal{Q}\left[U(x\vert y) C \gamma^5, C \gamma^5 D(x\vert y)\right]_{\alpha\beta,a' a} \\
        &\equiv \sum_{\gamma} \varepsilon_{abc} \varepsilon_{a'b'c'} \left(U(x\vert y) C\gamma^5\right)_{\alpha \gamma, bb'} \left(C\gamma^5 D(x\vert y)\right)_{\beta \gamma, cc'},
\end{align}
and where $U(x\vert y)$ and $D(x \vert y)$ denote the up and down quark propagator, respectively. In \autoref{eq:2pt_proton}, both contractions are of a type that shows up in all correlation functions with one baryon at source and sink; hence, translating both contractions in the distillation frameworks solves a whole class of correlation functions. The first contraction in \autoref{eq:2pt_proton} is of the trace-full type $\mathbb{T}_{ABC} = \trc{\trs{\mathcal{Q}[A, B]} C}$, where $A$, $B$ and $C$ can be an arbitrary product of a quark propagator and different $\Gamma$ structures. Analog to $\trc{A}$, $\trs{A}$ denotes a trace over all spin indices of $A$. In the language of distillation objects, contractions of this type can be expressed as 
\begin{align}\label{eq:trace-full}
    \mathbb{T}_{ABC}(\boldsymbol{p})_{\alpha \alpha'} &= \sum_{\boldsymbol{x},\boldsymbol{y}} e^{i \boldsymbol{p} \cdot (\boldsymbol{y} - \boldsymbol{x})} \mathcal{Q}\left[A(x\vert y),  B(x\vert y)\right]_{\beta \beta, a b} C_{\alpha \alpha', b a}(x \vert y) \\
    &= \mathcal{E}^{\ell n m}(t_x, \boldsymbol{p}) \mathcal{A}^{nn'}_{\beta\beta'}(t_x, t_y) \mathcal{B}^{mm'}_{\beta\beta'}(t_x, t_y) \mathcal{C}^{\ell \ell'}_{\alpha \alpha'}(t_x, t_y) \left[\mathcal{E}^{\ell' n' m'}(t_y, \boldsymbol{p})\right]^\dagger,
\end{align}
with an implicit sum over all repeated indices and the modified elemental given by 
\begin{equation}
    \mathcal{E}^{\ell n m}(t_x, \boldsymbol{p}) = \sum_{\boldsymbol{x}} \varepsilon_{abc} V_a^\ell(\boldsymbol{x}, t_x) V_b^n(\boldsymbol{x}, t_x) V_c^m(\boldsymbol{x}, t_x) e^{-i \boldsymbol{p} \cdot \boldsymbol{x}},
\end{equation}
which is of a similar form as used in \cite{Egerer_2019} and \cite{Lang_2017}, however, we are not including spin structures. The calligraphic versions of $A$, $B$, and $C$ denote the corresponding perambulators one obtains through the basis change from \autoref{eq:smeared_quark} multiplied with the same $\Gamma$-structure as its propagator. To reduce the cost of computing and storing the modified elementals, we can use the fact that the Levi-Civita tensor imposes its permutation behavior to the elementals. It is sufficient to compute all ordered tuples of $(\ell, n, m)$, i.e., $\ell < n < m$, and the remaining permutations of the tuple $\sigma(\ell, n, m)$ can be computed by
\begin{equation}
    \mathcal{E}^{\sigma(\ell, n, m)} = \text{sign}\left[\sigma(\ell, n, m)\right] \mathcal{E}^{\ell n m}.
\end{equation}
This reduces the required storage and runtime by the factor $R(N_d) = \frac{N_d (N_d-1) (N_d-2)}{6 N_d^3} = \frac16 - \frac{1}{2N_d} + \frac{1}{3N_d^2}$. 
The second contraction in \autoref{eq:2pt_proton} is of the trace-less type $\mathbb{L}_{ABC} = \trc{\mathcal{Q}[A,B] C}$, which does not have the spin trace over the quark contraction structure. For distillation, this contraction type is translated to
\begin{align}\label{eq:trace-less}
    \mathbb{L}_{ABC}(\boldsymbol{p})_{\alpha \alpha'} &= \sum_{\boldsymbol{x},\boldsymbol{y}} e^{i \boldsymbol{p} \cdot (\boldsymbol{y} - \boldsymbol{x})} \mathcal{Q}\left[A(x\vert y),  B(x\vert y)\right]_{\alpha \beta, a b} C_{\beta \alpha', b a}(x \vert y) \\
    &= \mathcal{E}^{\ell n m}(t_x, \boldsymbol{p}) \mathcal{A}^{nn'}_{\alpha\beta'}(t_x, t_y) \mathcal{B}^{mm'}_{\beta\beta'}(t_x, t_y) \mathcal{C}^{\ell \ell'}_{\beta \alpha'}(t_x, t_y) \left[\mathcal{E}^{\ell' n' m'}(t_y, \boldsymbol{p})\right]^\dagger.
\end{align}
In terms of trace-full and trace-less contractions, the nucleon two-point function reads
\begin{equation}
    C_{pp}(t, \boldsymbol{p}) = \left\langle\tr{P \left(\mathbb{T}_{UC\gamma^5,C\gamma^5 D,U}(\boldsymbol{p}) + \mathbb{L}_{UC\gamma^5,C\gamma^5 D,U}(\boldsymbol{p})\right)}\right\rangle,
\end{equation}
which only encompasses perambulators and modified elementals. 
These two tensors and the momentum insertion are sufficient to construct every $n$-point correlation function with at most one baryon. More details about the construction of $n$-point correlation functions are shown in \autoref{sec:contract}.

\newpage
\section{Computational Details}\label{sec:computational}

\subsection{Lattice details}
The ensembles used in this study have Iwasaki gauge action and $N_f = 2+1$ Möbius domain-wall fermion sea quarks \cite{brower2014mobius, Shamir_1993, Furman_1995, Blum_2016} generated by the RBC-UKQCD collaboration. \autoref{tab:ensembles} shows an overview of all the ensembles we used in the following. Ensembles 1, 3, and 4 are also used in \cite{RBC:2024fic}. The code for generating all ensembles is freely available in \cite{GPT}. All propagators of the light quarks are generated with the same mass as the light sea quarks. The autocorrelations between the individual configurations are small enough to consider them uncorrelated. For this reason, we do not perform any binning analysis. We use the single-elimination jackknife resampling for the statistical error estimate of the correlation functions and derived quantities. 
As a basis for our distillation setup, we use the eigenvectors derived from the 3-dimensional Laplace operator in \autoref{eq:laplace}, which are then used to compute the modified elementals, the perambulators, and momentum insertions. To improve the long-distance behavior of the distillation objects, we use Gaussian smearing applied to the link $U_i$ of the Laplacian in both spatial and temporal dimensions with the parameters $\rho = 0.1$ and $N = 30$. To maintain temporal locality during smearing, we first keep only a small window of three time slices $t-1, t, t+1$ non-zero and then apply the 4-dimensional smearing. Ultimately, we only preserve the data from the central time slice $t$. This process is iterated for each time slice across the entire configuration (see also \cite{Bruno:2023pde}). We fix the lattice spacing of all ensembles to the values obtained from physical pion mass ensembles with similar lattice spacing. In our case, ensemble 4, D, 1, 3, and C are fixed to the value of 48I, and the remaining are fixed to the values of 64I \cite{RBC:2024fic}. 

\begin{table}
\begin{tabular}{c|c|c|c|c|c|c|c}
    Ens-Id & $L^3 \times T \times L_s$ & $m_\pi / \text{MeV}$ & $m_K / \text{MeV}$ & $a^{-1}/\text{GeV}$ & $N_{\text{conf}}$ & $N_c$ & $m_\pi L$\\
    \hline
    4 & $24^3 \times 48 \times 24$ & $274.8(2.5)$ & $530.1(3.1)$ & $1.7312(28)$ & $94$ & $60$ & $3.8$\\ 
    D & $32^3 \times 64 \times 24$ & $274.8(2.5)$ & $530.1(3.1)$ & $1.7312(28)$ & $60$ & $60$ & $5.1$\\
    9 & $32^3 \times 64 \times 12$ & $278.9(4.9)$ & $531.2(4.9)$ & $2.3549(49)$ & $60$ & $60$ & $3.8$\\
    L & $64^3 \times 128 \times 24$ & $278.9(4.9)$ & $531.2(4.9)$ & $2.3549(49)$ & $20$ & $60$ & $7.6$\\
    1 & $32^3 \times 64 \times 24$ & $208.1(1.1)$ & $514.0(1.8)$ & $1.7312(28)$ & $34$ & $60$ & $3.8$\\ 
    3 & $32^3 \times 64 \times 24$ & $211.3(2.3)$ & $603.8(6.1)$ & $1.7312(28)$ & $34$ & $60$ & $3.8$ \\
    C & $64^3 \times 128 \times 24$ & $139.32(30)$ & $499.44(88)$ & $1.7312(28)$ & $25$ & $120$ & $5.2$\\ 
\end{tabular}
\caption{Overview over all ensembles used in this publication, where $N_c$ denotes the number of used eigenmodes of the Laplace operator. 
We use $N_{\text{conf}}$ independent configurations for the individual ensembles. The lattice spacings $a$ are taken from physical pion mass ensembles generated by the RBC-UKQCD collaborations, namely 48I and 64I\cite{RBC:2024fic}. The corresponding ensemble for ensembles 4, D, 1, 3, and C is 48I and 64I for the remaining. Also note that we use the same pion and kaon mass for ensembles, which differ only in volume. The pairs are (4, D), (9, L), and (48I, C), where the values are measured for the first ensemble of the pair.} 
\label{tab:ensembles}
\end{table}

\begin{figure}[h!]
    \begin{center}
        \includegraphics{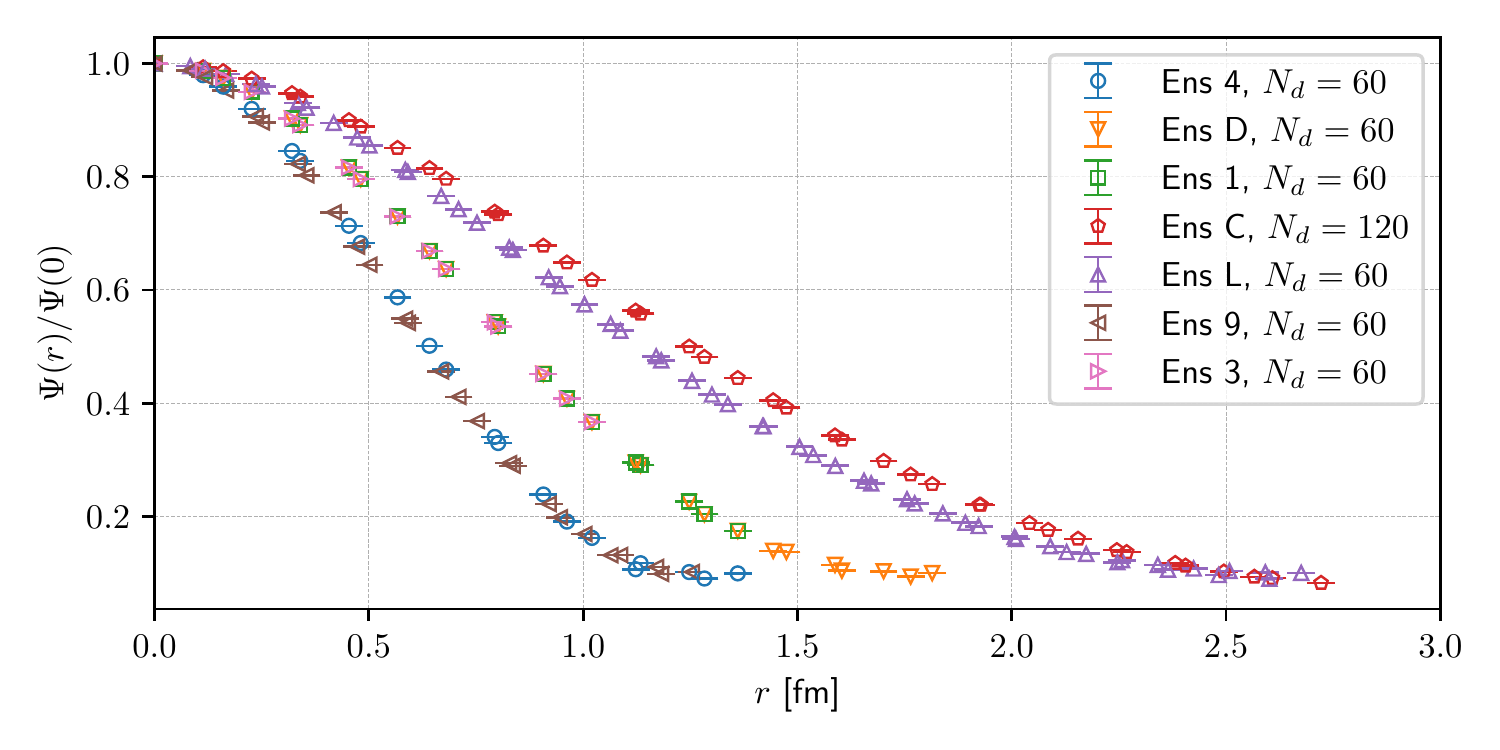}
    \end{center}
    \caption{The profile $\Psi(\boldsymbol{r}) = \sum_{\boldsymbol{x}, t} \sqrt{\trc{\Box(\boldsymbol{x}, \boldsymbol{x} + \boldsymbol{r}, t) \Box(\boldsymbol{x} + \boldsymbol{r}, \boldsymbol{x}, t)}}$ of all ensembles used in this study. The data points of Ensemble D and Ensemble 1 lie on each other, indicating that the pion mass does not contribute to the smearing of the distillation method. }
    \label{fig:profile}
\end{figure}

\subsection{All Mode Averaging (AMA)}
Throughout this study, we use All Mode Averaging \cite{DeGrand_2005, Bali_2010, Blum_2013, Shintani_2015}. Traditionally, the AMA is performed only on individual configuration, e.g., by computing the exact and sloppy propagator for a subset of source positions, and for a larger amount of source positions, we only compute the sloppy propagator. The sloppy computations are typically done with single precision and a less strict stopping condition for the inversion of the Dirac operator. For this study, we only compute the exact and sloppy propagators for a subset of configurations. The remaining configurations are only solved sloppily. 

\subsection{Automatic contractions}\label{sec:contract}

\begin{figure}[h!]
    \centering
    \begin{tikzpicture}[scale=1.0, decoration={
        markings,
        mark=at position 0.5 with {\arrow{>}}
    }]
        \draw[fill](-2.05,-0.25) rectangle +(0.1, 0.1);
        \draw[fill](-2, 0) circle [radius=0.05];
        \draw[fill](-2,0.2) circle [radius=0.05];
        
        \draw(-2, 0) ellipse [x radius = 0.13, y radius = 0.4] node [black, left=0.3] {$p$};

        \draw(-2, 0.9) circle [radius=0.05];
        \draw[fill](-2.05, 1.05) rectangle +(0.1, 0.1); 

        \draw(-2, 1) ellipse [x radius = 0.1, y radius = 0.3] node [black, left=0.2] {$\pi^-$};

        \draw(-2, 1.6) circle [radius=0.05];
        \draw[fill](-2.05,1.75) rectangle +(0.1, 0.1);

        \draw(-2, 1.7) ellipse [x radius = 0.1, y radius = 0.3] node [black, left=0.2] {$\pi^+$};

        \draw(1.95,-0.25) rectangle +(0.1, 0.1);
        \draw(2, 0) circle [radius=0.05];
        \draw(2, 0.2) circle [radius=0.05];
        \draw(2, 0) ellipse [x radius = 0.13, y radius = 0.4] node [black, right=0.2] {$p$};

        \draw[fill](2, 0.9) circle [radius=0.05];
        \draw(1.95, 1.05) rectangle +(0.1, 0.1);
        \draw(2, 1) ellipse [x radius = 0.1, y radius = 0.3] node [black, right=0.2] {$\pi^-$};

        \draw(1.95,1.75) rectangle +(0.1, 0.1);
        \draw[fill](2, 1.6) circle [radius=0.05];
        \draw(2, 1.7) ellipse [x radius = 0.1, y radius = 0.3] node [black, right=0.2] {$\pi^+$};

        %lines 
        \draw[tabblue, postaction={decorate}](-1.95, 0) to [out=0, in=180] (1.95, 0);

        \draw[tabblue, postaction={decorate}](-1.95, 0.2) to [out=0, in=0] (-1.95, 0.9);
        \draw[tabblue](-2, 0.95) to [out=90, in=270] (-2, 1.05);
        \draw[tabblue, postaction={decorate}](-1.95, 1.1) to [out=0, in=180] (1.95, 1.1);
        \draw[tabblue](2, 0.95) to [out=90, in=270] (2, 1.05);
        \draw[tabblue, postaction={decorate}](1.95, 0.9) to [out=180, in=180] (1.95, 0.2);

        \draw[tabblue, postaction={decorate}](-1.95, -0.2) to [out=0, in=180] (1.95, -0.2);

        \draw[tabred, postaction={decorate}](-1.95, 1.6) to [out=0, in=180] (1.95, 1.6);
        \draw[tabred](2, 1.65) to [out=90, in=270] (2, 1.75);
        \draw[tabred, postaction={decorate}](1.95, 1.8) to [out=180, in=0] (-1.95, 1.8);
        \draw[tabred](-2, 1.75) to [out=270, in=90] (-2, 1.65);

        %\draw[fill](3.95, -0.05) rectangle +(0.1, 0.1) node [black, right=0.0]{$ \sim \bar d$};
        %\draw(3.95,-0.35) rectangle +(0.1, 0.1) node [black, right=0.0]{$ \sim d$};
        %\draw[fill](4,-0.6) circle [radius=0.05] node [black, right=0.05]{$ \sim \bar u$};
        %\draw(4,-0.9) circle [radius=0.05] node [black, right=0.05]{$ \sim u$};
    \end{tikzpicture}
    \caption{Exemplary diagram illustrating the fact that every baryon and $n$ meson correlation function can be expressed as three (sequential) propagators connecting the baryonic fermions (blue lines), and the remaining mesonic quarks are part of a loop over multiple mesons (red lines).}
    \label{fig:example_sketch}
\end{figure}
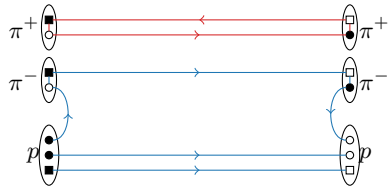

Throughout this study, we analyzed correlation functions $\langle \mathcal{O}_A(t,\boldsymbol{p}) \mathcal{O}_B^\dagger(0, \boldsymbol{q})\rangle$ of the five non-projected operators $\mathcal{O}_p$, $\mathcal{O}_{n\pi^+}$, $\mathcal{O}_{p\pi^0}$, $\mathcal{O}_{p\pi^0\pi^0}$ and $\mathcal{O}_{p\pi^+\pi^-}$. Each projected operator can be written as a linear combination of these elemental non-projected operators with different momenta. This results in only the $25$ elemental two-point correlation functions having to be properly contracted. However, a view on \autoref{tab:diagrams} suggests that the number of contractions grows factorially with the number of quarks, resulting in, e.g., 144 non-vanishing diagrams for $\left\langle\mathcal{O}_{p\pi^+\pi^-}(t, \boldsymbol{p}) \mathcal{O}^\dagger_{p\pi^+\pi^-}(t, \boldsymbol{p})\right\rangle$. To tackle this problem, we have implemented a code to automate the contraction of general nucleon-pion correlation functions with an arbitrary number of pions at source and sink. The main ideas for this automatization are illustrated in \autoref{fig:example_sketch}, where one contraction of the given example is shown. There are two main findings in the sketch. First, every quark in the baryon at the sink has to be connected to one anti-quark at the source baryon. This connection can be direct via a single propagator or sequentially over multiple pions (blue lines). The second finding is that every remaining pion, which is not part of a sequential propagator connecting baryonic quarks, must be part of a loop of propagators containing $N$ pions (red lines), equivalent to a trace of a sequential propagator.

\begin{table}[]
    \centering
    \begin{tabular}{c|ccccc}
         $\mathcal{O}_A/\mathcal{O}_B$ & $p$ & $n\pi^+$ & $p\pi^0$ & $p\pi^+\pi^-$ & $p\pi^0\pi^0$ \\
         \hline
        $p$ & $2$ & $4$ & $6$ & $12$ & $26$ \\
        $n\pi^+$ & $4$ & $12$ & $16$ & $36$ & $84$ \\
        $p\pi^0$ & $6$ & $16$ & $26$ & $60$ & $138$ \\
        $p\pi^+\pi^-$ & $12$ & $36$ & $60$ & $144$ & $372$ \\
        $p\pi^0\pi^0$ & $26$ & $84$ & $138$ & $372$ & $882$ \\
    \end{tabular}
    \caption{Number of non-vanishing diagrams for each correlation function $C_{AB}(t, \boldsymbol{p}, \boldsymbol{q}) = \left\langle \mathcal{O}_A(t,\boldsymbol{p}) \mathcal{O}^\dagger_B(0,\boldsymbol{q})\right\rangle$.}
    \label{tab:diagrams}
\end{table}

The sequential propagators have, in general, the form 
\begin{align}
    S_{\alpha \alpha', cc'}(y \vert x; \boldsymbol{p}_0, \boldsymbol{p}_1, \dots, \boldsymbol{p}_n) = \sum_{x_0, x_1, \dots x_n} G^{(0)}_{\alpha \beta_0, c b_0}(y \vert x_0) e^{-i \boldsymbol{p}_0 \cdot \boldsymbol{x}_0} \Gamma_0 G^{(1)}_{\beta_0 \beta_1, b_0 b_1}(x_0 \vert x_1) \dots G^{(n)}_{\beta_n \alpha', b_n c'}(x_n \vert y),
\end{align}
where $G^{(i)}$ are the general quark propagator with corresponding flavor and $\Gamma_i$ denote the $\Gamma$-structures of the included pions with momenta $\boldsymbol{p}_i$. With the use of perambulators $\mathcal{G}$ and momentum insertions $\mathcal{P}$, we can translate these sequential propagators to sequential perambulator by
\begin{align}
    \mathcal{S}(t_y, t_x; \boldsymbol{p}_0, \boldsymbol{p}_1, \dots, \boldsymbol{p}_n) = \mathcal{G}^{(0)}(t_y, t_{x_0}) \mathcal{P}(t_{x_0}, \boldsymbol{p}_0) \Gamma_0 \mathcal{G}^{(1)}(t_{x_0}, t_{x_1}) \dots \mathcal{G}^{(n)}(t_{x_n}, t_x).
\end{align}
The following algorithm then realizes the automatic contraction of the correlation: First, all fermionic fields are brought in a predefined order, and then Wick's theorem is applied to obtain the individual propagator. The next step is to find all sequential propagators connecting both baryons or forming a meson loop. After that, the remaining step is to classify the contraction as trace-full (see \autoref{eq:trace-full}) or trace-less (\autoref{eq:trace-less}) contraction, which can be translated to terms containing only distillation objects, i.e., perambulators, momentum insertions, and modified elementals. We compute all contractions contributing to the individual correlation functions using this algorithm. Due to the large number of different contraction topologies, we do not show the corresponding contraction diagrams. \autoref{fig:example_sketch} serves as an exemplary diagram from a large set of possible diagrams. The implementation of the automatic Wick contractions can be found in \cite{Hackl_AutoWick_2024}. In future work, we will consider methods such as the ones discussed in Ref.~\cite{Detmold:2012eu}.

\subsection{Wigner–Eckart theorem}
Prior to the isospin projection, there are in total five positive-parity ($\mathcal{O}_p$, $\mathcal{O}^P_{n\pi^+,G_{1g}}$, $\mathcal{O}^P_{p\pi^0, G_{1g}}$, $\mathcal{O}^S_{p\pi^+\pi^-, G_{1g}}$, $\mathcal{O}^S_{p\pi^0\pi^0, G_{1g}}$) and five negative-parity operator ($\mathcal{O}^-_p$, $\mathcal{O}^{(000)}_{n\pi^+,G_{1u}}$, $\mathcal{O}^{(000)}_{p\pi^0, G_{1u}}$, $\mathcal{O}^{(001)}_{n\pi^+,G_{1u}}$, $\mathcal{O}^{(001)}_{p\pi^0, G_{1u}}$) to consider, resulting in 25 correlation functions for each parity channel. This number can be reduced by using the so-called Wigner-Eckart theorem, which states that for a spherical tensor $T^{(k)}$ of rank $k$, there exists the relation
\begin{equation}
    \mel{jm}{T_q^{(k)}}{j'm'} = \braket{j'm'kq}{jm} \left\langle j \Big\vert \Big\vert T^{(k)} \Big\vert \Big\vert j'\right\rangle,
\end{equation}
where $\ket{jm}$ denotes the isospin eigenstates, $\braket{j'm'kq}{jm}$ the $SU(2)$ Clebsch-Gordan coefficients and $\left\langle j \vert \vert T^{(k)} \vert\vert j'\right\rangle$ the so-called reduced matrix elements. One example of a spherical tensor is the scattering operator $S$ of rank $0$, 
\begin{equation}
    \mel{jm}{S}{j'm'} = \delta_{mm'} S^j,
\end{equation}
with the reduced scattering amplitude $S^j$. Another useful tensor for our purpose is the pseudo-scalar operator $\mathcal P$ of rank $1$ with the entries 
\begin{align}
    \mathcal{P}^+(x) &= - \bar{d}(x) \gamma^5 u(x), \\
    \mathcal{P}^0(x) &= \frac{1}{\sqrt2} \left(\bar{u}(x) \gamma^5 u(x) - \bar{d}(x) \gamma^5 d(x)\right), \\
    \mathcal{P}^-(x) &= \bar{u}(x) \gamma^5 d(x). 
\end{align}
With these two spherical tensors, one can deduce the following relations between different two-point correlation functions: 
\begin{subequations}
\begin{align}
    C_{p,p\pi^0}(t) &= - \frac{1}{\sqrt{2}} C_{p,p\pi^+}(t) \\
    C_{p\pi^0,p\pi^0}(t) &= \frac{1}{\sqrt{2}} C_{n\pi^+,p\pi^0}(t) + C_{n\pi^+,n\pi^+}(t) \\
    C_{p,p\pi^0\pi^0}(t) &= - C_{p,p\pi^+\pi^-}(t) \\
    C_{p\pi^0\pi^0,p\pi^0\pi^0}(t) &= C_{p\pi^0\pi^0, p\pi^+ \pi^-}(t) + 2 C_{p\pi^+\pi^-,p\pi^+ \pi^-}(t) \\
    C_{p\pi^0, p \pi^+\pi^-}(t) &= \sqrt{2} C_{n\pi^+,p\pi^0\pi^0}(t) \\
    C_{p\pi^0,p \pi^0 \pi^0}(t) &= \frac{1}{\sqrt{2}} C_{n\pi^+,p\pi^0\pi^0}(t) + \sqrt{2} C_{n\pi^+ \to p\pi^+\pi^-}(t).
\end{align}
\end{subequations}
Further relation are created with the fact that $C_{AB}(t) = \left(C_{BA}(t)\right)^\star$ for general $A$ and $B$. These relations reduce the number of correlation functions we need to compute. Especially since correlation functions containing $\pi^0$ have significantly more contractions than those without. Besides the reduced computational cost, this theorem can also be used to cross-check implementations of the contraction code as it was introduced in \cite{Barca:2022pzc}.

\subsection{Fitting}\label{sec:fitting}
To extract the masses from the effective mass curves obtained using GEVP or single correlation functions, we make an excited state fit onto the effective mass curve using the fit form
\begin{equation}\label{eq:fit_form}
    f(am, A, aE^{\text{ex}}; t) = am + A e^{- (E^{\text{ex}} - m) t}.
\end{equation}
The optimal fit ranges $\tau_0$ to $\tau_1$ are obtained by comparing the excited state and plateau fit for different fit ranges, where we fix $\tau_1$ for the scanning of $\tau_0$. We choose $\tau_1$ to be in a region where the signal-to-noise ratio is still moderate to ensure the stability of the fit. Furthermore, we check whether the model $f$ can describe the data at the first time slice not contained in the fit, $\tau_0 - 1$. The measure we are using to determine the quality of the fit in this matter is the tension between the data at $m_{\text{eff}}(\tau_0 - 1)$ and the model estimate $f(am, A, aE^{\text{ex}}; \tau_0-1)$ defined by
\begin{equation}\label{eq:extrapol_check}
    \sigma_{\tau_0 - 1} = \frac{\sigma\left[m_{\text{eff}}(\tau_0 - 1) - f(am, A, aE^{\text{ex}}; \tau_0-1)\right]}{E\left[m_{\text{eff}}(\tau_0 - 1) - f(am, A, aE^{\text{ex}}; \tau_0-1)\right]},
\end{equation}
where $\sigma[X]$ and $E[X]$ denote the standard deviation and expected values of $X$. We aim our fits to have a tension of $\sigma_{\tau_0 - 1}< 2$. In the following, we denote $\sigma_{\tau_0 - 1}$ as extrapolation check.
Since we have a small number of independent configurations for some ensembles, we perform uncorrelated fits for these ensembles. Because the $p$-value is not a good quality measure for uncorrelated fits, we use instead the reduced chi-square values $\chi^2/\text{dof}$ as an additional measure for the performance of the fit.
In summary, we consider $\chi^2/\text{dof}$, the tension $\sigma_{\tau_0-1}$, and the relation to other excited states and plateau fits as measures for the quality of the fits. In the end, one of the fits that fulfills the criteria we choose as our preferred fit for the mass estimates. The variance estimates are done using fits of the jackknife samples obtained from the single-elimination jackknife method.  

\subsection{Normalization scheme of GEVP eigenvectors}\label{sec:fixed-point}
The square matrix consisting of the eigenvectors of the GEVP for given $t$ and $t_0$, i.e., $\boldsymbol{V} = (V_1, \dots, V_n)$, can be normalized in two different ways. First, each eigenvector $V_n$ can be normalized individually by introducing a normalization factor $\alpha_n$. The second way stems from the fact that we can multiply each individual operator $\mathcal{O}_m$ by a constant factor $\beta_m$ without changing its physical properties. This change of $\beta_m$ is equivalent to normalizing the individual rows of $\boldsymbol{V}$. Each normalization in one direction destroys the normalization of the perpendicular direction. However, one can show that a repeated change of normalization in both directions converges to a transformed matrix satisfying $\sum_n \vert V_{nm}(t, t_0) \vert^2 = 1$ and $\sum_m \vert V_{nm}(t, t_0) \vert^2 = 1$ for fixed values of $t$ and $t_0$. Throughout this work, we will use this scheme to normalize all GEVP eigenvectors. 

\section{Positve Parity Results}\label{sec:positive-parity}
Our spectroscopy analysis contains a GEVP analysis with each nucleon operator and two multi-hadron operators for every parity channel and ensemble. In the GEVP, we use a constant time distance between both correlation matrices of $\Delta t = a$, which enables us to make an excited state fit on each effective mass curve of the fit form in \autoref{eq:fit_form}.

\subsection{Individual Ensemble GEVP}

\begin{figure}
    \centering
    \includegraphics[width=0.8\textwidth]{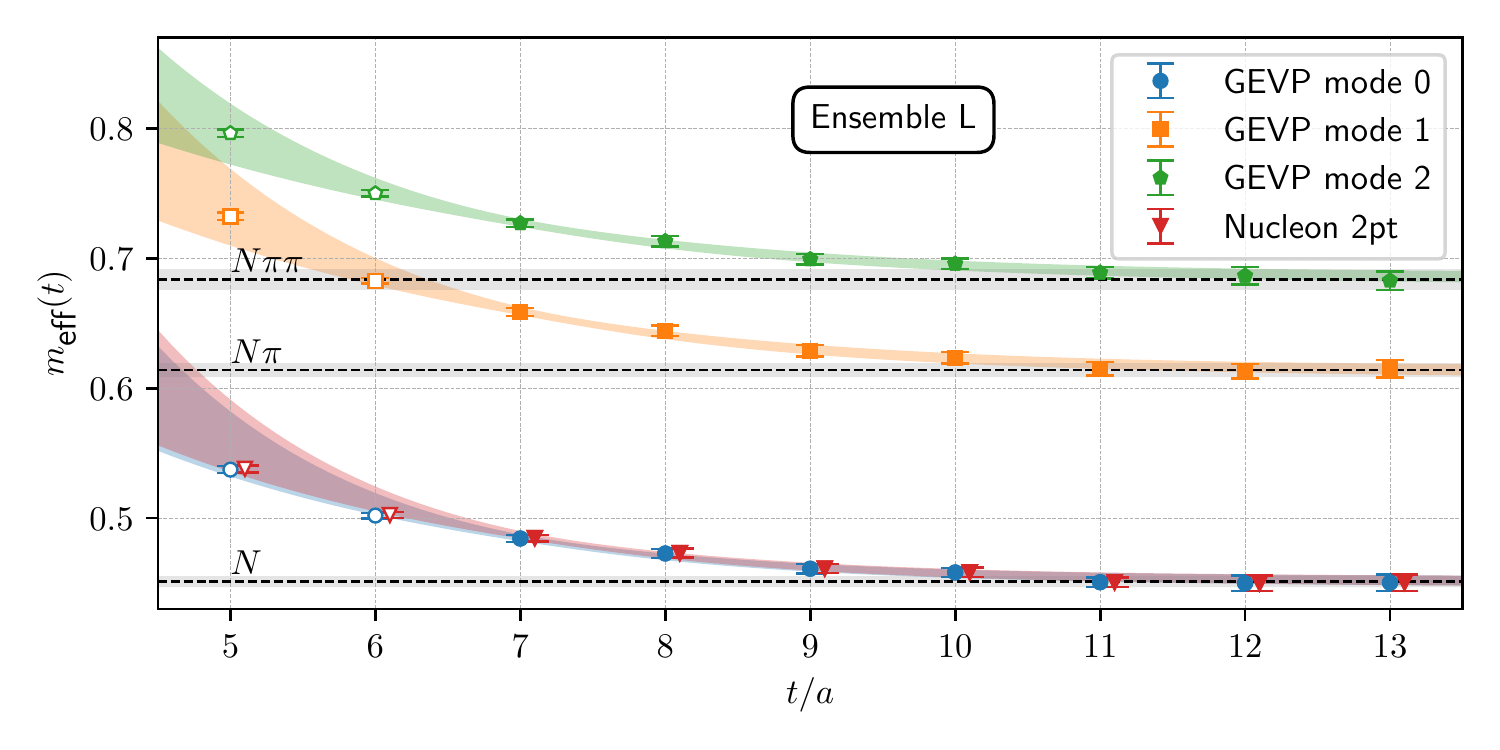}
    \caption{Plot of the effective energy of the three GEVP modes compared with the effective mass of the nucleon 2-point function of ensemble L. The bands depicted in the same color as the data points show the results of an excited state fit according to \autoref{eq:fit_form}. The fit contains all filled data points. The dashed line labeled by $N$ shows the central value of the nucleon mass obtained from the excited state fit of GEVP mode 0. The remaining lines, labeled by $N\pi$ and $N\pi\pi$, show the energies of the non-interacting $N\pi$ state in p-wave and $N\pi\pi$ state in s-wave. The shaded region around the lines represents the $1\sigma$ error estimate of the energy.}
    \label{fig:gevp-L}
\end{figure}

\begin{figure}
    \centering
    \includegraphics[width=0.8\linewidth]{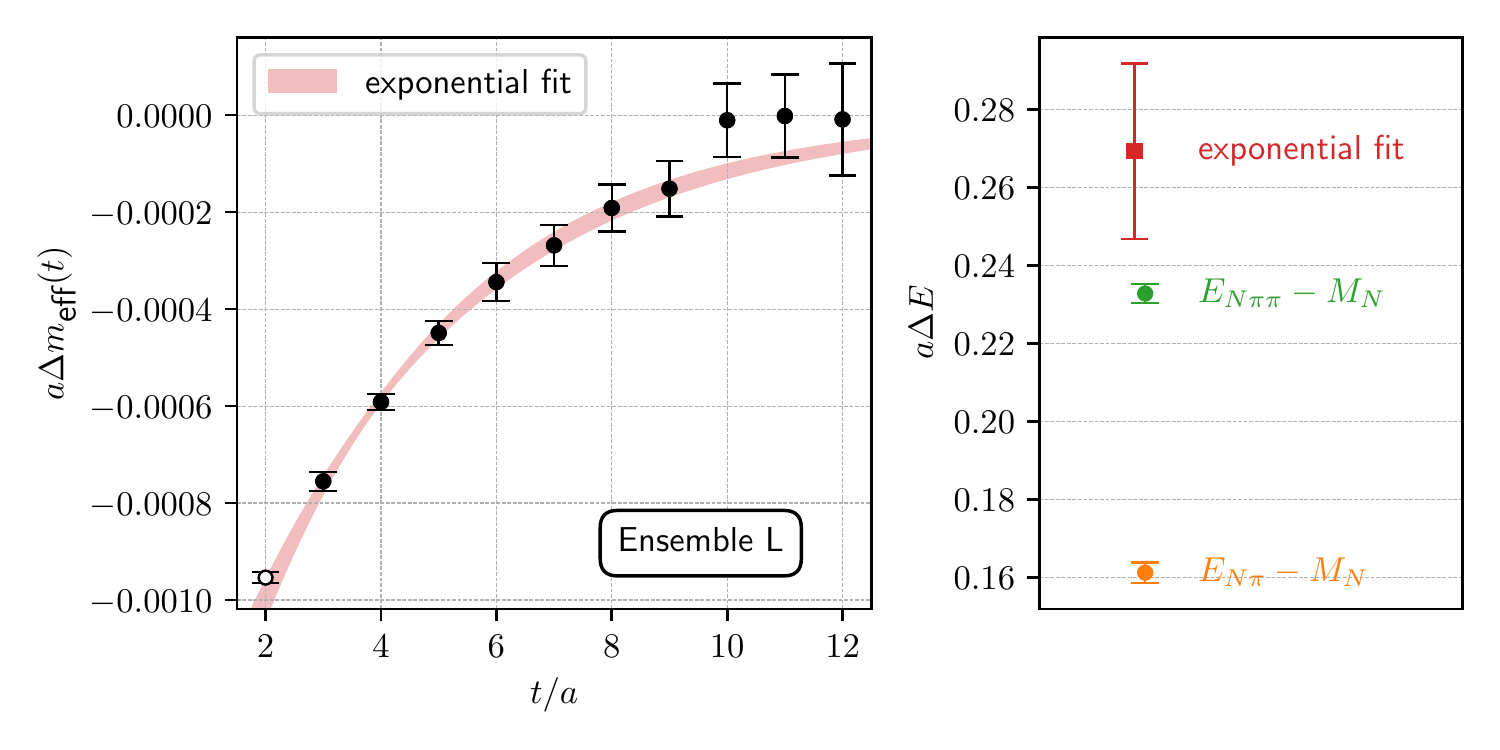}
    \caption{The data points of the left plot show the difference $\Delta m_{\rm eff}$ of effective masses between GEVP mode 0 and the nucleon 2-point function of ensemble L defined in Eq.~\ref{eqn:deltameff}. The red band represents the fit result of an exponential fit over all filled data points with the fit form $f(t; A, \Delta E) = A e^{-\Delta E t}$. The estimate for the energy gap $a \Delta E$ obtained from the fit is plotted in the right plot in comparison with the energy gaps obtained from the GEVP, namely $E_{N\pi\pi} - M_N$ and $E_{N\pi} - M_N$. }
    \label{fig:diff-L}
\end{figure}

\begin{figure}
    \centering
    \includegraphics[width=0.8\textwidth]{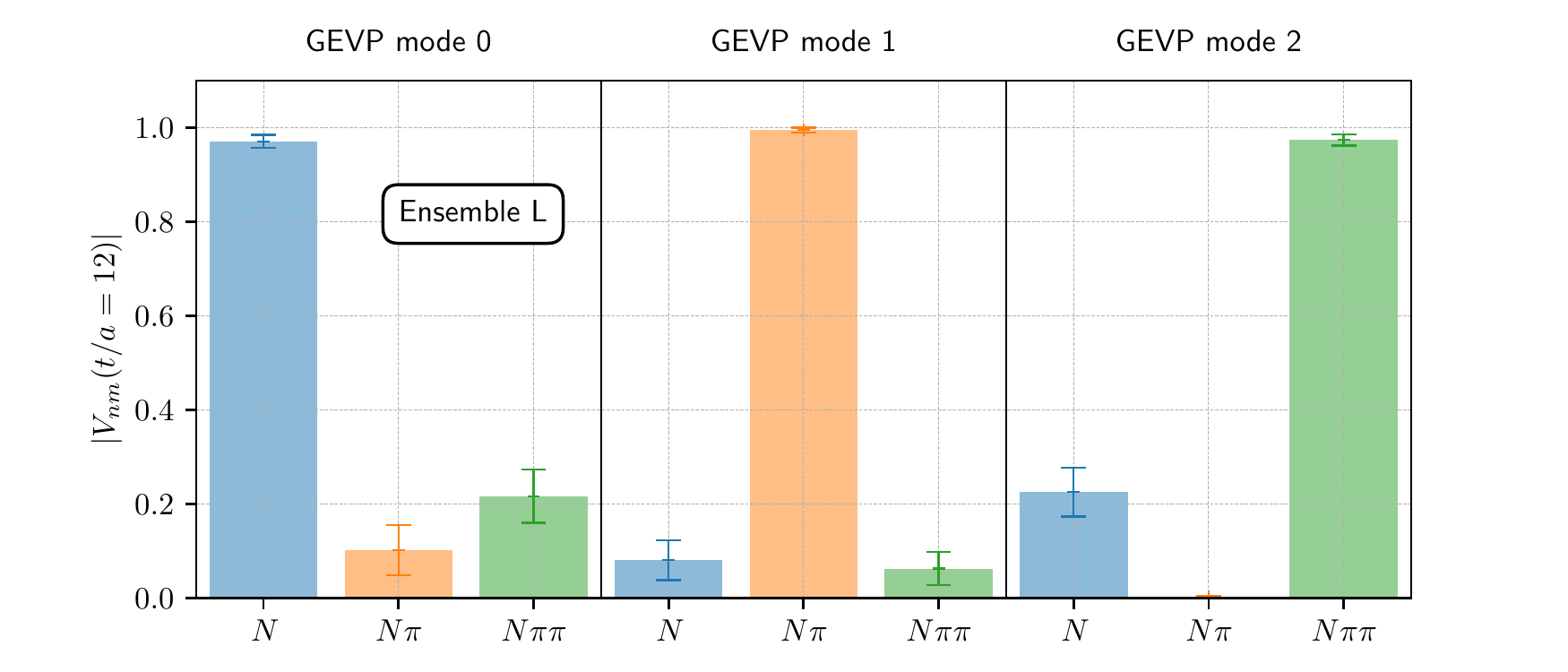}
    \caption{GEVP eigenvectors at $t/a = 12$ for ensemble L. The values are obtained using the fixed-point normalization scheme shown in \autoref{sec:fixed-point}.}
    \label{fig:evec-L}
\end{figure}

In the positive parity channel, we use the nucleon (\autoref{eq:nuc_op}), the p-wave nucleon-pion (\autoref{eq:npi_pos_parity}), and the s-wave nucleon-pion-pion (\autoref{eq:npp_swave}) interpolating operators, creating a $3\times 3$ generalized eigenvalue problem. The analysis is done for all ensembles of \autoref{tab:ensembles}. In the following, we will focus on the analysis of ensemble L since most of the statements are also true for the remaining ensembles.

The effective masses $a m_{\text{eff}}(t)$ of all GEVP modes are shown in \autoref{fig:gevp-L}. We also show the effective mass curve of the nucleon two-point function in red for comparison. All effective mass curves converge to a stable plateau, where the signal-to-noise ratio is still small, and the eigenmodes are separated. All curves are fitted against the fit form introduced in \autoref{eq:fit_form}, where the filled data points are included in the fit. The model estimate obtained from the fit is shown by the band colored in the color of the corresponding data points, which shows the $1\sigma$ confidence region of the estimate. The criteria used to determine the fit ranges are discussed in \autoref{sec:fitting}. The dashed line labeled by $N$ in \autoref{fig:gevp-L} denotes the mean nucleon mass taken from the GEVP mode 0 fit. The remaining lines show the dimensionless energies we expect from non-interacting particles. For $N\pi$ we get the energy $aE_{N\pi} = a(\sqrt{m_N^2 + p^2} + \sqrt{m_\pi^2 + p^2})$, where $m_N$ is the expectation value for the nucleon mass, the pion mass $m_\pi$ and $a$ are taken from \autoref{tab:ensembles} and $p$ denotes the momentum of the particle defined by $a p = \frac{2\pi}{L}$. The non-interacting $N\pi\pi$ energy is $a E_{N\pi\pi} = m_N + 2 m_\pi$. An important result shown in \autoref{fig:gevp-L} is that there is only a marginal difference between the effective mass of the GEVP mode 0 and the nucleon two-point function. This difference,
\begin{equation}\label{eqn:deltameff}
    a \Delta m_{\text{eff}}(t) = a m_{\text{eff}}^{\text{GEVP0}}(t) - a m_{\text{eff}}^{\text{2pt}}(t),
\end{equation}
is plotted in \autoref{fig:diff-L}, showing that the relative difference is on the per mil level. For the nucleon two-point function, we can approximate the effective mass curve by
\begin{equation}
    a m_{\text{eff}}^{\text{2pt}}(t) = a m_N + A_{N\pi} e^{-(E_{N\pi} - m_{N})t} + A_{N\pi\pi} e^{-(E_{N\pi\pi} - m_{N})t} + A_{N^\star} e^{-(E_{N^\star} - m_{N})t},
\end{equation}
whereas the effective mass of GEVP mode 0 has no overlap with $N\pi$ and $N\pi\pi$, since they are basis states of the GEVP, resulting in an approximate behavior of
\begin{equation}
    a m_{\text{eff}}^{\text{GEVP0}}(t) = a m_N + A'_{N^\star} e^{-(E_{N^\star} - m_N) t}.
\end{equation}
Combining both approximations yields the fit ansatz
\begin{equation}
    a \Delta m_{\text{eff}}(t) \propto e^{-(E_{X} - m_N) t},
\end{equation}
with $X$ being $N\pi$ or $N\pi\pi$ with the assumption that only $N\pi$ or $N\pi\pi$ is dominant. Fitting this ansatz yields the result shown by the red band in \autoref{fig:diff-L}. The criteria for finding a valid fit range are the same as in the previous fits of the effective masses. On the right subplot of \autoref{fig:diff-L}, we compare the energy gap $\Delta E = E_X - m_N$ we get from the fit with the energy gaps obtained from the GEVP mass estimates. We find that the estimate obtained from the single exponential fit coincides within a $2 \sigma$ margin with the gap $E_{N\pi\pi} - M_N$, indicating that $N\pi\pi$ is the dominant multi-hadronic contribution in the nucleon two-point function of this ensemble concerning our set of operators.   
A look at \autoref{fig:evec-L} confirms this observation. \autoref{fig:evec-L} shows the GEVP eigenvectors at $t/a = 8$ for ensemble L, normalized by the normalization scheme introduced in \autoref{sec:fixed-point}. The result for GEVP mode 0 shows that the dominant part of this state is created by the nucleon operator, GEVP mode 1 is dominated by contributions of the $N\pi$ operator, and GEVP mode 2 by the ones of the $N\pi\pi$ operator. We further obtain that GEVP mode 2, corresponding to the $N\pi\pi$ state, has a larger nucleon operator contribution than the $N\pi$ (GEVP mode 1) state. This agrees with the result from the exponential fit of $a \Delta m_{\text{eff}}$. The still considerable excited state contamination of GEVP mode 0 shows that there are additional states to consider.  We will study a larger operator basis in future work.

For the other ensembles, we obtain similar results. We can generally distinguish between the different eigenmodes, and all effective masses converge to the nucleon mass or close to the energy of the corresponding non-interacting particles (see \autoref{fig:all_gevp}). Furthermore, for all ensembles, we observe only a marginal difference between the 0th mode of the GEVP and the nucleon 2-point function. The exponential fit on this difference yields energy gaps between a $2\sigma$ margin to one of the gaps obtained from GEVP (see \autoref{fig:all_diff}). For the eigenvectors of the remaining ensembles, we also obtained that the dominant contribution coincides with the state we expect from its energy. We also find that, in general, the state with the second largest contribution from the nucleon operator matches with the state whose energy gap is closest to the estimate for the energy gap obtained from the single exponential fit of $\Delta m_{\text{eff}}$. 

In summary, we obtain that the $N\pi$ and $N\pi\pi$ contributions are negligible for the nucleon two-point function and subsequently, the nucleon mass estimate as it was already suggested in \cite{B_r_2015, B_r_2018} using chiral perturbation theory. For the $N\pi$ and $N\pi\pi$ states, we obtain new operators, which have no overlap with the ground state and the other multi-hadronic state contained in the GEVP operator set.

\begin{figure}
    \centering
    \includegraphics[width=\linewidth]{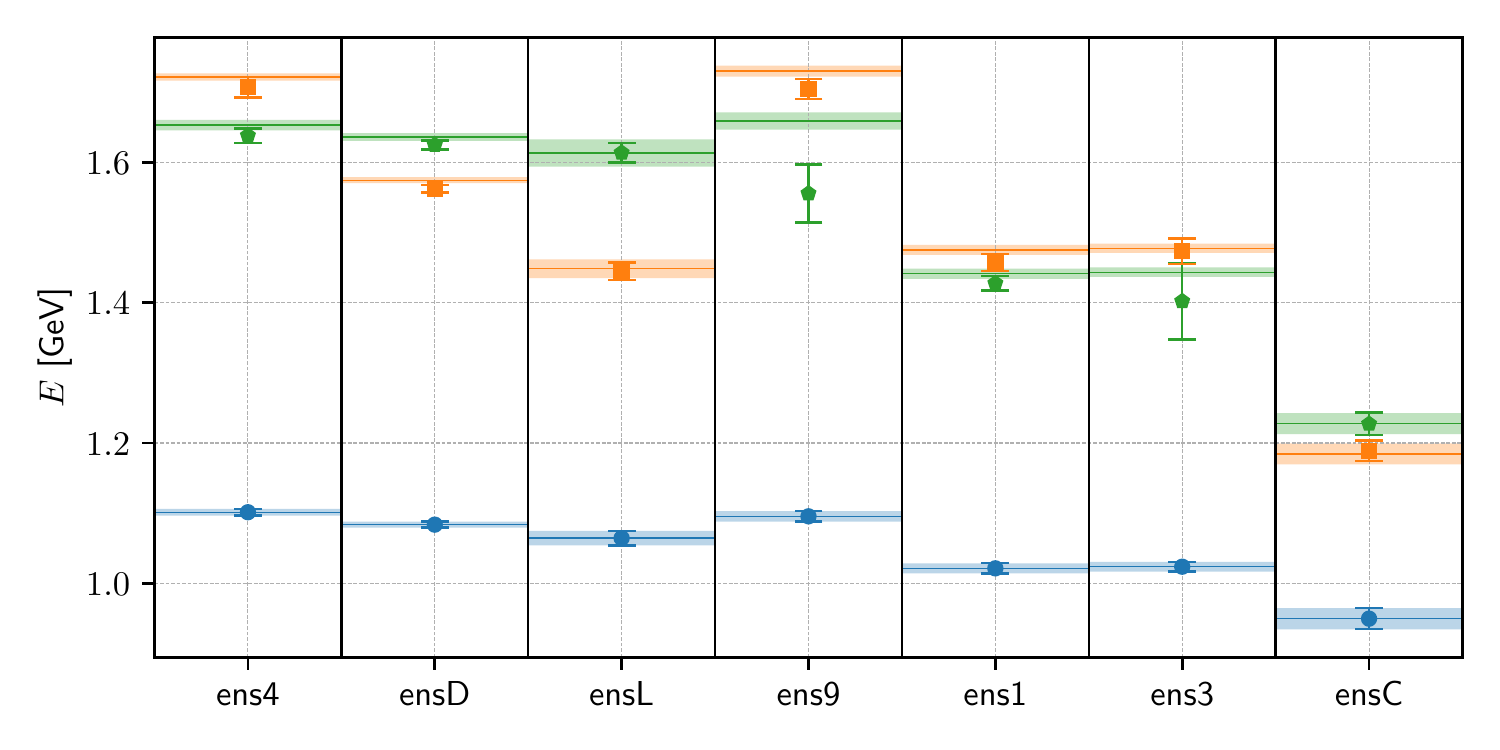}
    \caption{Overview of the energies of the individual ensembles. The data points describe the states' energy obtained from the individual positive parity GEVP analyses. The colors represent the corresponding states, i.e., blue represents the nucleon, orange is the $N\pi$ state, and green is the $N\pi\pi$ state. The blue lines correspond to the central value of the nucleon, and the remaining lines represent the non-interacting energies of the multi-hadronic states. The shaded regions around the lines represent the $1\sigma$ error estimate of the non-interacting energies.}
    \label{fig:volume_dependency}
\end{figure}

\autoref{fig:volume_dependency} gives an overview over the energy of all $N$, $N\pi$, and $N\pi\pi$ state for the different ensembles. For most ensembles, the energies of the multi-hadronic states coincide with the non-interacting energies of the particles. The only exception is ensemble 9, where a 2$\sigma$ tension is seen.  
The fit results for all ensembles are summarized in \autoref{tab:results} with a short explanation in \autoref{app:fit_result}. 

\subsection{Finite volume correction}
For the positive parity sector of this study, we are using the finite volume correction obtained from relativistic $SU(2)_f$ baryon chiral perturbation theory (B$\chi$PT) \cite{Ali_Khan_2004, Procura_2004}. The finite volume correction is given by
\begin{equation}
    m_N(L) - m_N(\infty) = \Delta_a(L) + \Delta_b(L),
\end{equation}
where $\Delta_a(L)$ and $\Delta_b(L)$ denote contributions of order $\mathcal{O}(p^3)$ and $\mathcal{O}(p^4)$, respectively. The explicit form of $\Delta_a(L)$ is
\begin{equation}
    \Delta_a(L) = \frac{3 g_A^2 m_0 m_\pi^2}{16 \pi^2 f_\pi^2} \int_{0}^{\infty} \text{d} x \sum_{\boldsymbol{n} \neq \boldsymbol{0}} K_0\left( L \vert \boldsymbol{n} \vert \sqrt{m_0^2 x^2 + m_\pi^2 (1-x)}\right),
\end{equation}
where $K_0$ denotes the modified Bessel function and $\boldsymbol{n}$ are integer vectors (cf \cite[Equation (15)]{Ali_Khan_2004}). The constants $g_A$, $f_\pi$, and the nucleon mass $m_0$ are taken from the chiral limit. The contribution $\Delta_b$  (cf. \cite[Equation (23)]{Ali_Khan_2004}) is 
\begin{equation}
    \Delta_b(L) = \frac{3 m_\pi^4}{4\pi^2 f_\pi^2} \sum_{\boldsymbol{n} \neq \boldsymbol{0}} \left[(2 c_1 - c_3) \frac{K_1(L \vert \boldsymbol{n} \vert m_\pi)}{L \vert \boldsymbol{n} \vert m_\pi} + c_2 \frac{K_2(L \vert \boldsymbol{n} \vert m_\pi)}{(L \vert \boldsymbol{n} \vert m_\pi)^2}\right],
\end{equation}
where new coupling constants, $c_1$, $c_2$ and $c_3$, are introduced. 

\begin{figure}
    \centering
    \includegraphics[width=0.8\linewidth]{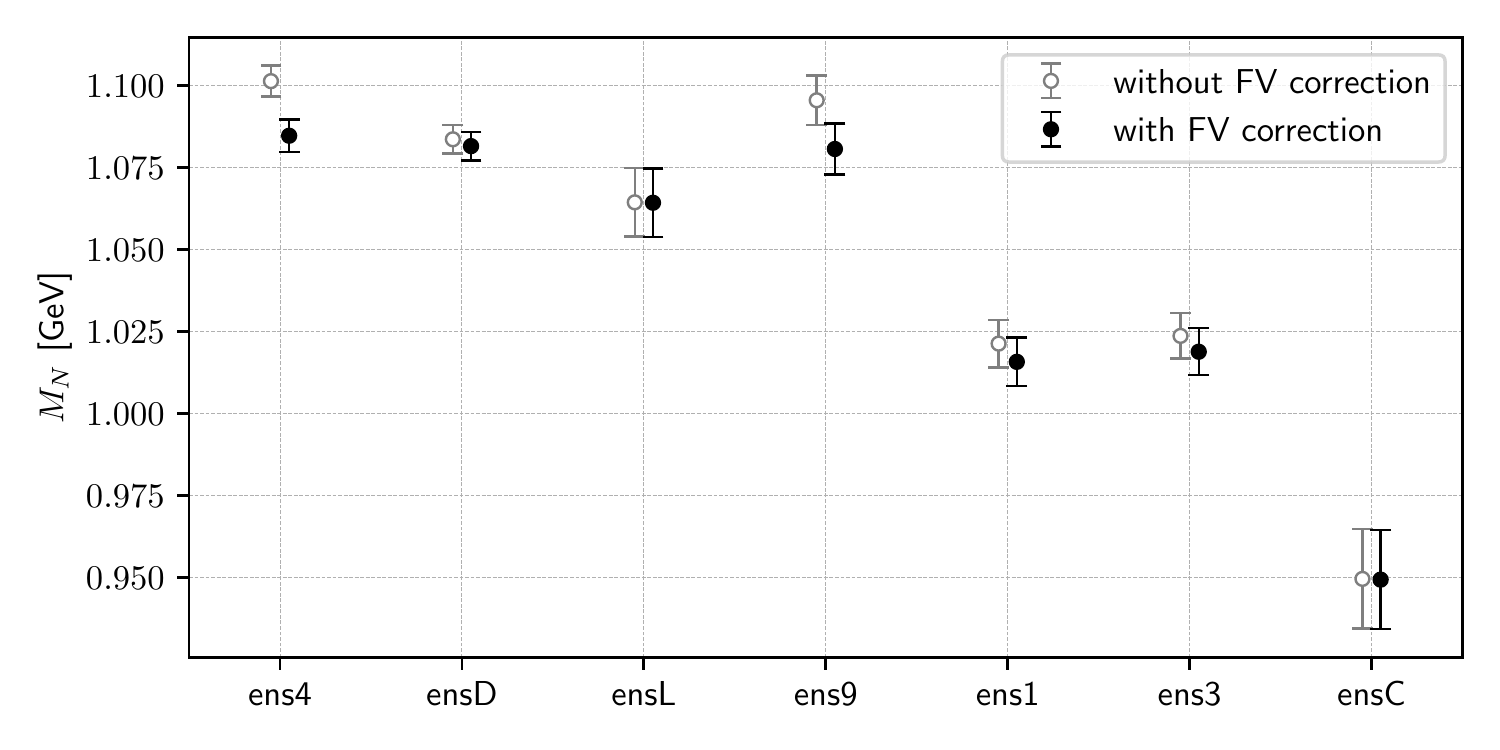}
    \caption{Finite volume (FV) correction of the nucleon masses for the different ensembles. The empty and filled data points show the values before and after the finite volume correction.}
    \label{fig:mass-correction}
\end{figure}

The explicit values for the finite-volume correction are taken from \cite{Bali_2013}, where we use the same preselected values for $g_A$, $f_\pi$, $c_2$ and $c_3$, and for the remaining parameter, we take the fit results obtained from a combined fit of nucleon mass and $\sigma$-term data. We take the first weighted average in Table B.4 of \cite{Bali_2013}. The preselection for $g_A$ and $f_\pi$ are the physical values $g_A = 1.256$ and $f_\pi = 92.4\text{ MeV}$. For $c_2$ and $c_3$ the values are taking according to \cite{meißner2005quarkmassdependencebaryon, Bali_2013} to be $c_2 = 3.3(2) \text{ GeV}^{-1}$ and $c_3 = - 4.7(1.3) \text{ GeV}^{-1}$. The remaining values obtained from the fit are $r_0 m_0 = 2.26(4)$ and $c_1/r_0 = -0.34(3)$, with $r_0 = 2.58(4) \text{ GeV}^{-1}$. Using error propagation, we obtain the values in physical units, which reads $m_0 = 0.89(3) \text{ GeV}$ and $c_1 = -0.78(8) \text{ GeV}^{-1}$. \autoref{fig:mass-correction} shows the effect of the finite volume correction on the nucleon mass estimate of the individual ensembles. The difference between ens4 and ensD is its volume; hence, after correcting for the finite volume, we expect the two data points to align, which is validated by \autoref{fig:mass-correction}. To underline this statement by some numbers: before the finite-volume correction, the data points for ens4 and ensD had an approximate tension of $2.8 \sigma$. After the correction, the tension drops to $0.5 \sigma$, hinting that our choice for finite volume correction is valid. We obtain a similar observation concerning ensL and ens9, which also only differ in the volume. The statistical uncertainties of the parameters are propagated by creating fake jackknife samples with the wanted Gaussian distribution.  

\subsection{Continuum extrapolation of the nucleon mass}

The extrapolation to the continuum and physical pion and kaon masses is done by a model averaging using the Akaike information criterion\cite{Akaike1998, Akaike74} $\text{AIC} = 2k + \chi^2$. Using this criterion, we obtain a model average for the variable $\beta$, which is
\begin{equation}\label{eq:model_average}
    \bar{\beta} = \sum_{\mathcal{M}} P(\mathcal{M}) \beta_{\mathcal{M}},
\end{equation}
where $\mathcal{M}$ denote the specific model and $\beta_{\mathcal{M}}$ the parameter obtained from this model $\mathcal{M}$. $P(\mathcal{M})$ indicates the probability of the model $\mathcal{M}$ and reads 
\begin{equation}
    P(\mathcal{M}) = \frac{e^{-\text{AIC}_{\mathcal{M}}}}{\sum_{\mathcal{M}'} e^{-\text{AIC}_{\mathcal{M'}}}}.
\end{equation}
We compute the statistical error using jackknife resampling and for the systematic error of the model averaging, we obtain
\begin{equation}
    \text{Var}(\beta)_{\text{sys}} = \sum_{\mathcal{M}} P(\mathcal{M}) (\beta_{\mathcal{M}} - \bar{\beta})^2. 
\end{equation}
We chose our models to obtain the nucleon mass dependency of the lattice spacing $a$, the pion mass $m_\pi$, and the kaon mass $m_K$. We expect discretization errors of $\mathcal{O}(a^2)$ for domain-wall fermions. Hence, we describe the lattice spacing dependency by a quadratic term, $c_1 a^2$. Since we are interested in the nucleon mass for physical pion and kaon mass, $m_\pi^0$ and $m_K^0$, we embed the pion and kaon terms such that the constant term of our model $M_N$ resembles the nucleon mass at the physical point. We chose the kaon dependency to be linear, quadratic, or nonexistent. This is combined with two different types of pion contributions. One is inspired by chiral perturbation theory \cite{Ali_Khan_2004}, which starts with a quadratic term. For consistency, we also include the cubic terms in our set of model functions. Besides the chiral perturbation theory ansätze, we also use linear models inspired by \cite{walkerloud2008newlessonsnucleonmass}.

\begin{figure}
    \centering
    \includegraphics[width=\linewidth]{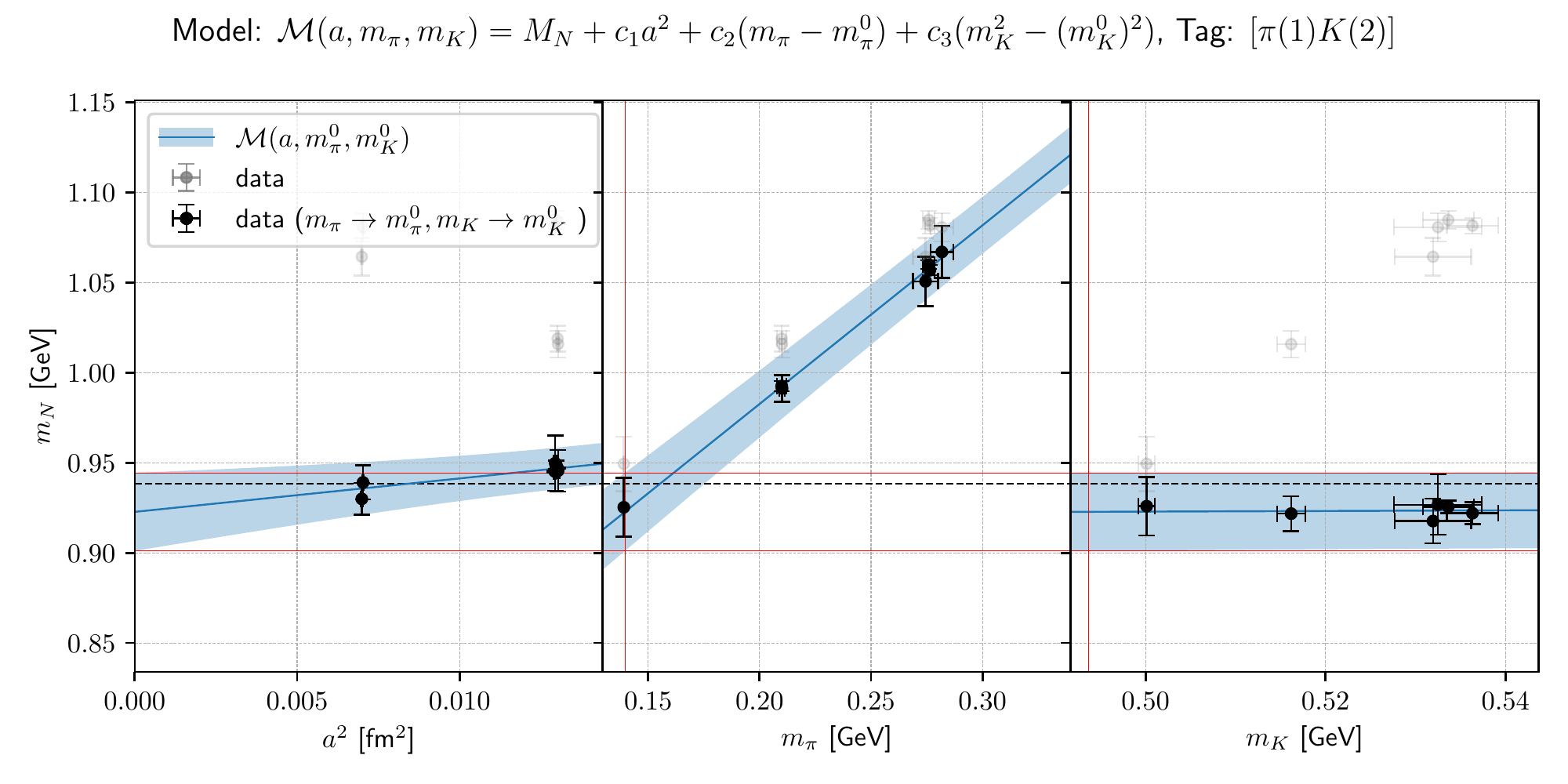}
    \caption{Plot of the fit results of the physical point extrapolation. Each subplot shows the fitted model's dependency for one of the variables $a$, $m_\pi$ or $m_K$. The original data shown by the transparent gray points is projected to the physical values of the perpendicular parameters shown by the black points. The red horizontal lines denote the upper and lower limit of the $1\sigma$ confidence interval of the nucleon mass, and the vertical lines represent the physical pion and kaon mass.}
    \label{fig:limit_example}
\end{figure}

In total, we consider nine models for our model averaging, all summarized in \autoref{tab:overview_continuum_model}. The results of the linear pion-mass model with a quadratic kaon-mass term are shown in \autoref{fig:limit_example}, which consists of three plots, each illustrating the dependency of the nucleon mass for one of the three parameters. The two parameters not considered in the plot are set to the physical point. We also project the original data point (grey transparent data points) to the physical point (filled black data points). The red horizontal lines in the plots show the $1\sigma$ confidence interval of the nucleon mass, whereas the vertical lines show the physical pion and kaon mass. The plots for the remaining models are shown in \autoref{fig:lin_model_summary}, \autoref{fig:quad_model_summary}, and \autoref{fig:quad_cub_model_summary}. In this project, we have not measured the individual configuration's pion and kaon masses or the lattice spacing. To include the statistical uncertainty of these quantities, we have created fake data with the same statistical properties. This fake data is then treated like the real data, and its uncertainties are propagated using the single elimination jackknife.

\begin{table}[h!]
    \centering
    \begin{tabular}{l|l}
        Tag & Model function $\mathcal{M}(a, m_\pi, m_K)$ \\
        \hline
        $\pi(1)K(0)$ & $M_N + c_0 a^2 + c_1 (m_\pi - m^0_\pi)$ \\
        $\pi(1)K(1)$ & $M_N + c_0 a^2 + c_1 (m_\pi - m^0_\pi) + c_2 (m_K - m_K^0)$ \\
        $\pi(1)K(2)$ & $M_N + c_0 a^2 + c_1 (m_\pi - m^0_\pi) + c_2 (m_K^2 - (m_K^0)^2)$ \\
        \hline
        $\pi(2)K(0)$ & $M_N + c_0 a^2 + c_1 (m_\pi^2 - (m_\pi^0)^2)$ \\
        $\pi(2)K(1)$ & $M_N + c_0 a^2 + c_1 (m_\pi^2 - (m_\pi^0)^2) + c_2 (m_K - m_K^0)$ \\
        $\pi(2)K(2)$ & $M_N + c_0 a^2 + c_1 (m_\pi^2 - (m_\pi^0)^2) + c_2 (m_K^2 - (m_K^0)^2)$ \\
        \hline 
        $\pi(2,3)K(0)$ & $M_N + c_0 a^2 + c_1 (m_\pi^2 - (m_\pi^0)^2) + c_2 (m_\pi^3 - (m_\pi^0)^3)$ \\
        $\pi(2,3)K(1)$ & $M_N + c_0 a^2 + c_1 (m_\pi^2 - (m_\pi^0)^2) + c_2 (m_K - m_K^0) + c_3 (m_\pi^3 - (m_\pi^0)^3)$ \\
        $\pi(2,3)K(2)$ & $M_N + c_0 a^2 + c_1 (m_\pi^2 - (m_\pi^0)^2) + c_2 (m_K^2 - (m_K^0)^2) + c_3 (m_\pi^3 - (m_\pi^0)^3)$ \\
    \end{tabular}
    \caption{Overview of all models we use for the continuum extrapolation.}
    \label{tab:overview_continuum_model}
\end{table}

\begin{figure}
    \centering
    \includegraphics[width=\linewidth]{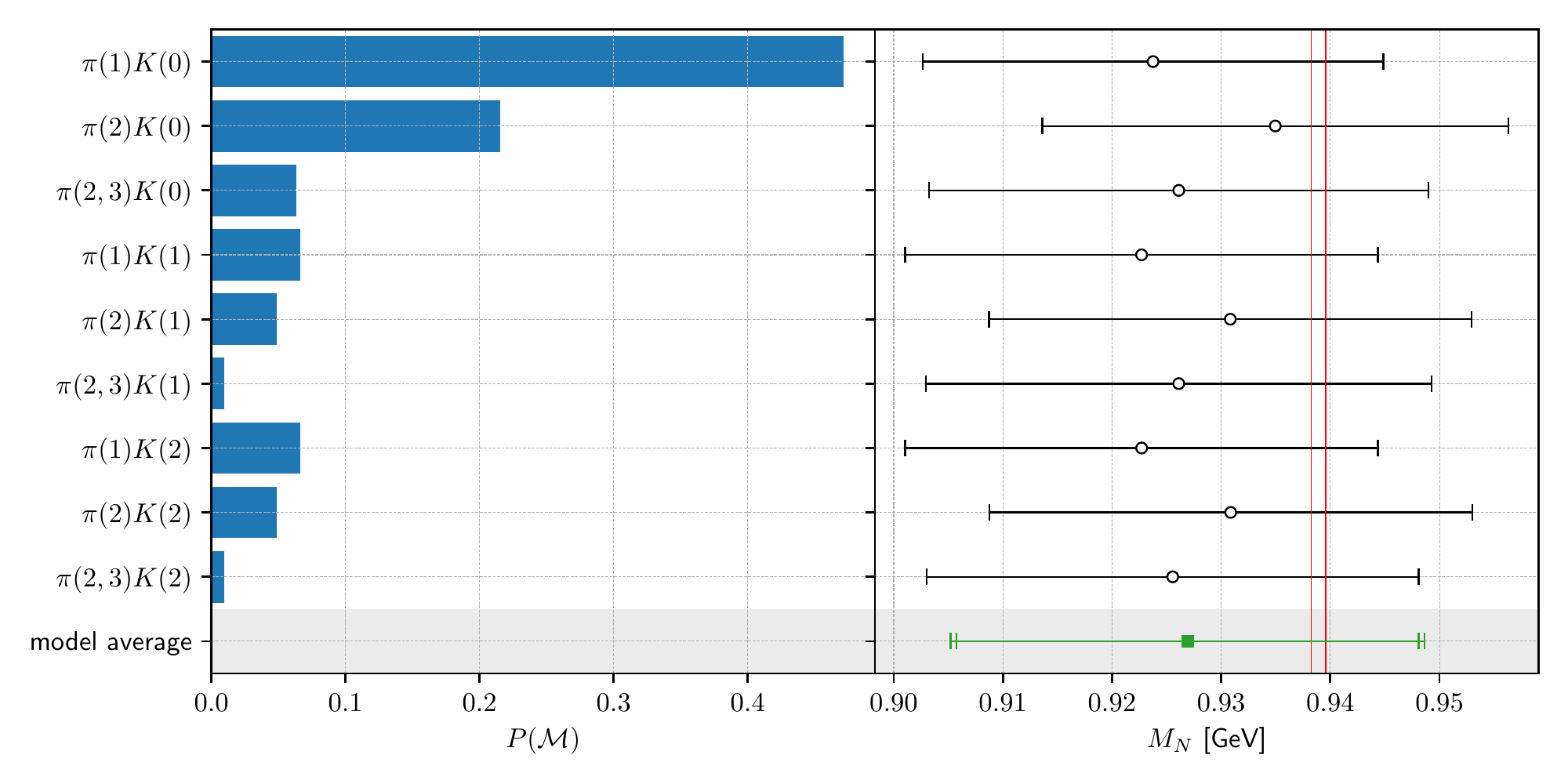}
    \caption{On the left are the probabilities of the different models using the Akaike information criterion. On the right, an overview of the nucleon mass estimates of the different models and the model average computed using \autoref{eq:model_average}. Both vertical lines show the experimental values of the proton and neutron mass. The tags of the models are defined in \autoref{tab:overview_continuum_model}. }
    \label{fig:limit_summary}
\end{figure}

The fit results for all models are shown in \autoref{tab:continuum_result}, where we can conclude that we have no kaon dependency and the cubic pion term is not constrained enough by our data, which is not surprising considering that we only have three pion masses.  The slope of the linear model is also consistent with values obtained by the ruler approximation \cite{walkerloud2008newlessonsnucleonmass}. Concerning the probabilities $P(\mathcal{M})$ (see \autoref{fig:limit_summary}), we see a slight preference for the linear model. From the model averaging, we obtain the final value for the nucleon mass to be
\begin{equation}
    M_N = 0.927(21)_{\text{stat}}(05)_{\text{sys}} \text{GeV},
\end{equation}
which is within the $1\sigma$ margin of the physical nucleon mass. 

\begin{table}[h!]
    \centering
    \renewcommand{\arraystretch}{1.5}
    \begin{tabular}{l|cccccccc}
       Tag  & $M_N / \text{GeV}$ & $c_0/ \text{GeV}^3$ & $c_1$ & $c_2^{(1)}$ & $c_2^{(2)} \cdot \text{GeV}$ &  & $\chi^2/\text{dof}$ & p-value \\
       \hline
       $\pi(1)K(0)$ & $0.924(21)$ & $0.073(52)$ & $0.987(83)$ & \textemdash & \textemdash & \textemdash & $0.89/4$ & $0.93$ \\
       $\pi(1)K(1)$ & $0.923(22)$ & $0.073(52)$ & $0.990(83)$ & $0.02(11)$ & \textemdash & \textemdash & $0.85/3$ & $0.84$ \\
       $\pi(1)K(2)$ & $0.923(22)$ & $0.072(53)$ & $0.991(83)$ & \textemdash & $0.017(98)$ & \textemdash & $0.85/3$ & $0.84$ \\
       \hline
       \hline
       Tag  & $M_N / \text{GeV}$ & $c_0/ \text{GeV}^3$ & $c_1 \cdot \text{GeV}$ & $c_2^{(1)}$ & $c_2^{(2)} \cdot \text{GeV}$ & $c_3 \cdot \text{GeV}^2$ & $\chi^2/\text{dof}$ & p-value \\
       \hline
       $\pi(2)K(0)$ & $0.935(21)$ & $0.081(55)$ & $2.14(18)$ & \textemdash & \textemdash & \textemdash & $1.67/4$ & $0.79$ \\
       $\pi(2)K(1)$ & $0.931(22)$ & $0.079(55)$ & $2.17(19)$ & $0.07(11)$ & \textemdash & \textemdash & $1.15/3$ & $0.76$ \\
       $\pi(2)K(2)$ & $0.931(22)$ & $0.079(55)$ & $2.17(19)$ & \textemdash & $0.065(98)$ & \textemdash & $1.15/3$ & $0.76$ \\
       \hline
       $\pi(2,3)K(0)$ & $0.926(23)$ & $0.073(52)$ & $4.4(2.3)$ & \textemdash & \textemdash & $-6.4(6.6)$ & $0.89/3$ & $0.83$ \\
       $\pi(2,3)K(1)$ & $0.926(23)$ & $0.074(53)$ & $3.8(2.9)$ & $0.04(12)$ & \textemdash & $-4.6(8.1)$ & $0.77/2$ & $0.68$ \\
       $\pi(2,3)K(2)$ & $0.926(23)$ & $0.074(53)$ & $4.0(2.8)$ & \textemdash & $0.03(11)$ & $-5.2(7.9)$ & $0.77/2$ & $0.68$ \\
    \end{tabular}
    \caption{Results of the different physical point extrapolation. The tags are defined in \autoref{tab:overview_continuum_model}. $c_2^{(1)}$ and $c_2^{(2)}$ denote the prefactor of the linear and quadratic kaon-mass term, respectively.}
    \label{tab:continuum_result}
\end{table}

\newpage

\section{Negative Parity Results}\label{sec:negative-parity}

The spectral analysis of the negative parity sector consists of a GEVP analysis of the nucleon operator (cf. \autoref{eq:nuc_op}) with negative parity projection, the nucleon-pion operator, with both particles at rest (cf. \autoref{eq:npi_swave}) and with a back-to-back momentum configuration (cf. \autoref{eq:npi_neg_parity}).

\begin{figure}
    \centering 
    \includegraphics[width=0.8\textwidth]{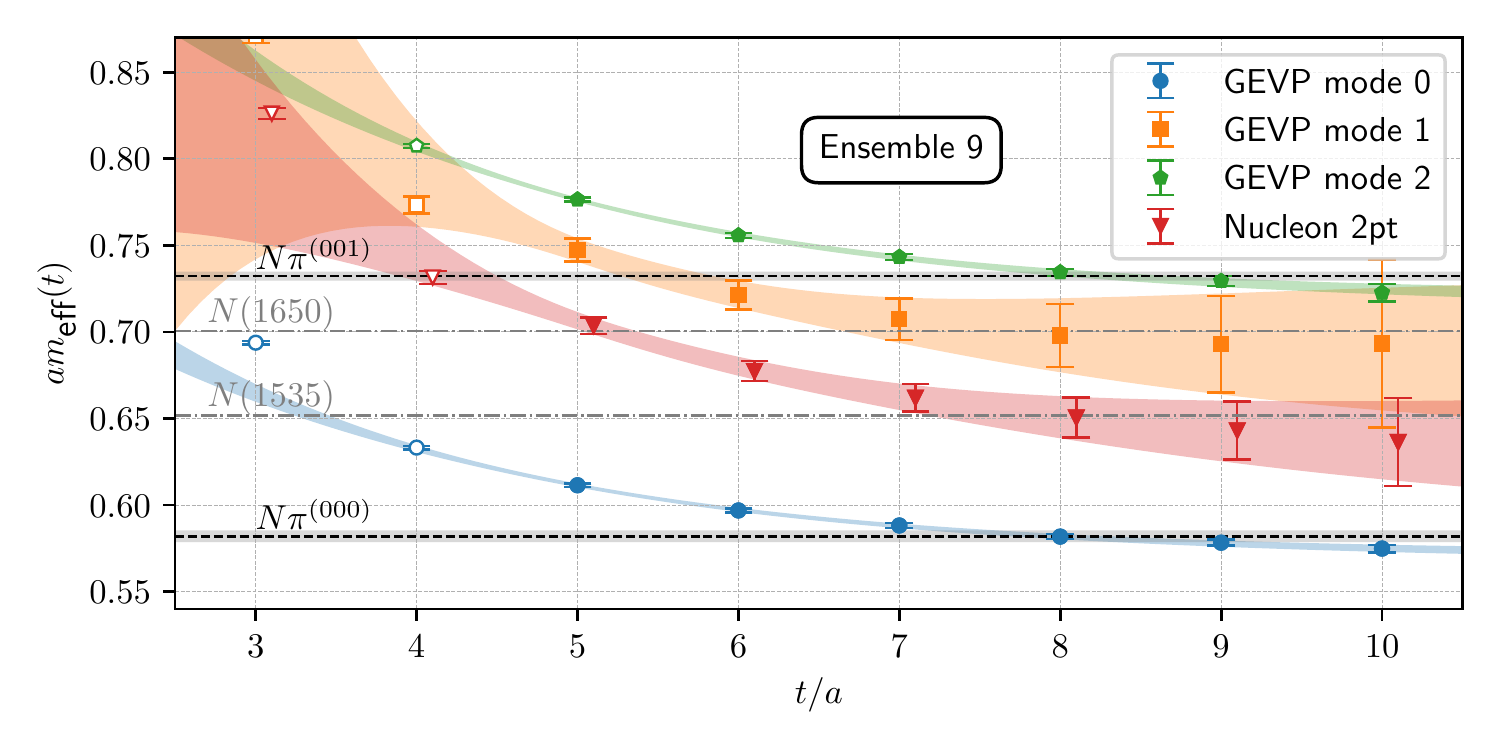}
    \caption{Effective mass curves of the GEVP eigenstates (GEVP mode 0, GEVP mode 1, GEVP mode 2) in the negative parity sector of ensemble 9 in comparison with the corresponding nucleon 2-point effective mass curve. The black dashed lines represent the non-interacting energies of the nucleon-pion state with both particles at rest, $N\pi^{(000)}$, and with back-to-back momentum of the magnitude $\vert \boldsymbol p\vert = \frac{2\pi}{aL}$ labeled by $N\pi^{(001)}$. The grey regions around the lines represent the $1\sigma$ error estimate of the energy. The dash-dot lines represent the physical masses of $N(1535)$ and $N(1650)$.}
    \label{fig:ngevp-9}
\end{figure}

\begin{figure}
    \includegraphics[width=0.8\textwidth]{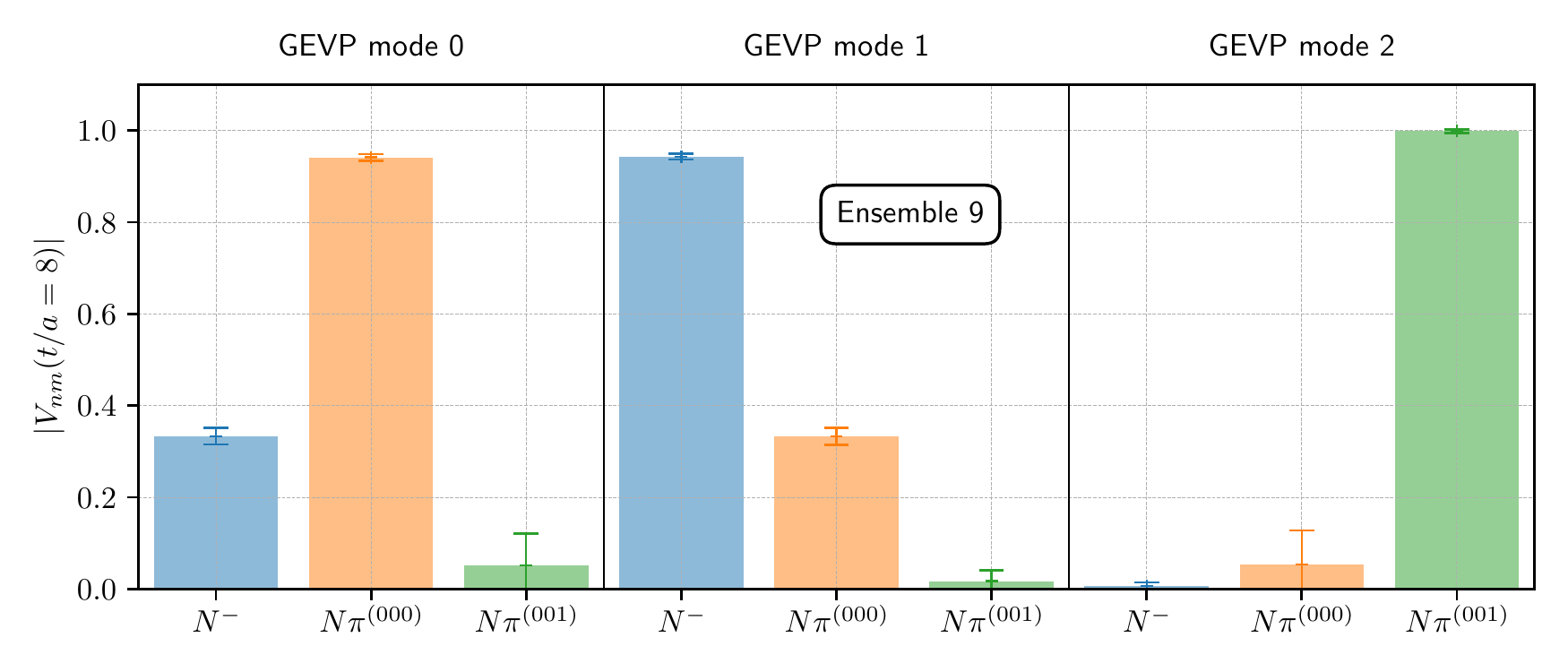}
    \caption{Negative parity GEVP eigenvectors at $t/a = 8$ of ensembles 9 renormalized using the scheme introduced in \autoref{sec:fixed-point}.}
    \label{fig:ngevp-9-evec}
\end{figure}

\autoref{fig:ngevp-9} shows the effective energies of the GEVP for ensemble 9 compared with the negative parity nucleon 2-point effective mass displayed by the red data points. Similar to the positive parity channel, we perform an excited state fit on all effective energies. The black dashed lines denote the energies of the non-interacting $N\pi$ states, where the nucleon mass estimate is obtained from the positive parity channel fits. The gray dash-dot lines show the energies of the two lightest negative parity nucleon states $N(1535)$ and $N(1650)$ obtained from the particle data group \cite{ParticleDataGroup:2024cfk}. The corresponding eigenvectors at time slice $t/a = 8$ are shown in \autoref{fig:ngevp-9-evec}, normalized according to \autoref{sec:fixed-point}.


The GEVP modes with the most significant overlap to the multi-hadronic states have a strong signal. However, the nucleon 2-point correlation function suffers a strong signal-to-noise degradation. We find a similar behavior for the remaining ensembles, shown in \autoref{fig:ngevp-all} with its eigenvectors \autoref{fig:ngevp-all-evec}. The nucleon 2-point function for most ensembles is even worse than for ensemble 9. One possible explanation is that we do not include sufficiently many high modes in our distillation smearing. This claim is supported by the fact that the two ensembles with the clearest signal, 4 and 9, have the most narrow smearing profile as shown in \autoref{fig:profile}.

However, for all of our ensembles, we find a clear signal for the multi-hadronic states. Furthermore, in contrast to the positive parity sector, the nucleon 2-point function significantly overlaps with the nucleon-pion states. Nevertheless, making a more sophisticated analysis in this sector would require a more considerable amount of Laplace eigenmodes in the distillation setup or the usage of a different strategy to include the distillation modes like, for example, the one introduced in \cite{Knechtli:2022bji}. As for the positive parity channel, all fit results of the individual excited state fits done in the GEVP analysis can be found in \autoref{tab:results} with a further explanation in \autoref{app:fit_result}. 

\section{Discussion and Conclusions}\label{sec:conclusion}

We can draw different conclusions for the different parity channels considered in this paper. We have obtained clear and stable results for the GEVP analysis in the positive parity sector throughout the ensembles listed in \autoref{tab:ensembles} by using a nucleon, a nucleon-pion, and, for the first time, a nucleon-pion-pion operator as operator set. We get a clear signal for the ground state for most of the ensembles, enabling a mass extraction with relative errors around 0.3 \% to 2 \%. For the multi-hadronic excited states, we find energies in the region of the non-interacting energy (cf. \autoref{fig:volume_dependency}). The only minor exception is ensemble 9, for which both the $N\pi\pi$ and $N\pi$ energy lie approximately $2\sigma$ below the non-interacting energy. The global significance of this fluctuation is, however, marginal at this point.  We also reproduce the chiral perturbation theory expectation of the negligibility of $N\pi$ and $N\pi\pi$ state contributions to the nucleon 2-point function \cite{B_r_2015}. We observed this behavior by finding only a marginal difference between the nucleon 2-point function's effective mass and the one of the ground-state GEVP mode. We could even quantify the extent of the multi-hadronic contributions by considering the difference between the nucleon 2-point function and the ground-state GEVP effective mass. Furthermore, the energy gap estimate obtained from an exponential fit on this difference coincides within a $2\sigma$ margin with the energy gap obtained from the GEVP between the ground state and the state with the second largest overlap with the nucleon operator. This overlap is obtained from the eigenvectors of the GEVP.

Following the GEVP analysis for the different ensembles, we performed finite volume corrections to the nucleon masses using $SU(2)_f$ baryon chiral perturbation theory. To test the validity of these corrections, we compared the values of pairs of ensembles that differ only in volume, namely Ensemble 4 and D and Ensemble L and 9. We find both pairs to be in agreement within the error margin.

Furthermore, we performed a physical point extrapolation of the nucleon mass using nine models with linear, quadratic, and quadratic + cubic pion mass dependencies. All models had reasonable $\chi^2/{\text{dof}}$ values. Hence, we performed a model average using the Akaike information criterion. The final value for the nucleon mass is within the error margin of the experimental proton and nucleon mass. We did not consider any isospin breaking and QED corrections since our final nucleon mass estimate has uncertainties too large to be sensitive to these corrections. The main goal of this study was to explore to which degree our distillation setup, which was optimized for our g-2 program, can be used for baryon physics.  We observe that our setup is also well suited for this task, and we will build on this setup in future work.

In the negative parity sector, we performed similar GEVP analyses with a nucleon operator and two nucleon-pion operators as the GEVP operator set. The main conclusion is that the number of distillation modes is too small for most ensembles to perform sophisticated analyses. We found for all ensembles that the multi-hadronic states can be resolved quite as well as for the positive parity case. However, the negative parity nucleon 2-point function suffers a substantial signal-to-noise degradation. The claim that more distillation modes are needed is supported by the fact that the ensembles with the sharpest profiles offer the best results for the negative parity nucleon state.  We are currently exploring a combination of distillation and all-to-all
methods similar to what we have done for the muon g-2 \cite{Bruno:2019nzm,RBC:2024fic}.

\section{Acknowledgments}
We want to thank our collaborators in the RBC and UKQCD collaborations for the fruitful discussions. We are grateful to Aaron Meyer for valuable comments regarding an earlier version of the manuscript.  The research of A.H. is supported by a scholarship from the Hans-Böckler-Foundation.  The contractions were performed on the local QPace3 and QPace4 clusters at the University of Regensburg.  The distillation data was generated using the following compute grants.
The authors gratefully acknowledge the Gauss Centre for Supercomputing
e.V. (www.gauss-centre.eu) for funding this project by providing
computing time on the GCS Supercomputer JUWELS at Jülich
Supercomputing Centre (JSC).  We acknowledge the EuroHPC Joint
Undertaking for awarding this project access to the EuroHPC
supercomputer LUMI, hosted by CSC (Finland) and the LUMI consortium
through a EuroHPC Extreme Scale Access call.  An award of
computer time was provided by the ASCR Leadership Computing Challenge
(ALCC) and Innovative and Novel Computational Impact on Theory and
Experiment (INCITE) programs. This research used resources of the
Argonne Leadership Computing Facility, which is a DOE Office of
Science User Facility supported under contract DE-AC02-06CH11357. This
research also used resources of the Oak Ridge Leadership Computing
Facility, which is a DOE Office of Science User Facility supported
under Contract DE-AC05-00OR22725.   We
gratefully acknowledge disk and tape storage provided by USQCD and by
the University of Regensburg with support from the DFG.
The contractions were performed using the automatic contractor "AutoWick" \cite{Hackl_AutoWick_2024}.
The distillation data was generated using GPT
\cite{GPT} which uses Grid \cite{GRID,Boyle:2016lbp,Yamaguchi:2022feu}
for performance portability.

\newpage

\bibliographystyle{apsrev4-1}
\bibliography{main}

\newpage
\appendix

\section{Interpolating operator construction}\label{sec:interpolating_operator_construction}

For interpolating operators with back-to-back momentum, the projection to the irreducible representations $G_{1g/u}$ is not a priori obvious. For a given spin $\alpha$ and momentum $\boldsymbol{p}$ the non-projected nucleon-pion operators read
\begin{equation}\label{eq:non-projected-Np-op}
    \mathcal{O}_{N\pi }(t; \boldsymbol p)_\alpha = \mathcal{O}_N(\boldsymbol{p})_\alpha \mathcal{O}_\pi(-\boldsymbol{p}).
\end{equation}
In this study, we use only finite momenta of the form $(0,0,1)$, i.e., there is only one non-zero component with $p_i = \frac{2\pi}{aL}$. Hence, for this type of momentum, we have six different momenta, $\pm\hat{p}_x, \pm\hat{p}_y$, and $\pm\hat{p}_z$. Combined with the spin degrees of freedom, we obtain a $24$-dimensional carrier space with a matrix representation $\Gamma$. This representation is reducible and consists of the direct sum of the irreducible representation $G_{1,u/g}$ and $H_{u/g}$. The projection onto the irreducible representation $\Lambda$ can be realized by the projection operator defined by
\begin{equation}
    P^\Lambda = \frac{\text{dim}(V)}{\vert G \vert} \sum_{g \in O_h^d} \Gamma(g) \chi_\Lambda(g),
\end{equation}
where $\text{dim}(V)$ denotes the dimension of the carrier space, $\vert G \vert$ the number of elements in the double cover of the full octahedral group $O_h^d$, and $\chi_\Lambda(g)$ represents the character of the irreducible representation $\Lambda$ for the group element $g$. The eigenvectors $V^\Lambda_{i}$ of the projection operator with eigenvalues $1$ produce linear combinations of operators that transform in the irreducible representation $\Lambda$. To further determine the spin direction of the newly formed operators, we compute directly the eigenvectors of $\Gamma(R_z) P^\Lambda$, where $\Gamma(R_z)$ denotes the representation of a rotation of $90^\circ$ around the $z$ axis. The projected operators are then
\begin{equation}
    \mathcal O^\Lambda_{N\pi}(t)_\alpha = \sum_{\beta} \sum_{\boldsymbol q, \vert \boldsymbol q \vert = \frac{2\pi}{aL}} V^{\Lambda,\alpha}_{\beta, q} \mathcal{O}_{N\pi}(t; \boldsymbol{q})_\beta,
\end{equation}
with $V^{\Lambda,\beta}_{\alpha, q}$ being the elements of the eigenvectors. Since the operators for all odd and even spins are identical, we denote even and odd spins by $\uparrow$ and $\downarrow$, respectively. \\
For the positive parity operator, we obtain then
\begin{equation}
\begin{aligned}
    \mathcal{O}^P_{N\pi, G_{1g}}(\boldsymbol 0, t)_{\uparrow} = \mathcal{O}_{N\pi}(t, \hat{p}_x)_{\downarrow} - \mathcal{O}_{N\pi}(t, -\hat{p}_x)_\downarrow  - i \mathcal{O}_{N\pi}(t, \hat{p}_y)_\downarrow + i \mathcal{O}_{N\pi}(t,-\hat{p}_y)_\downarrow +\mathcal{O}_{N\pi}(t, \hat{p}_z)_\uparrow - \mathcal{O}_{N\pi}(t, -\hat{p}_z)_\uparrow \\
    \mathcal{O}^P_{N\pi, G_{1g}}(\boldsymbol 0, t)_{\downarrow} = \mathcal{O}_{N\pi}(t, \hat{p}_x)_{\uparrow} - \mathcal{O}_{N\pi}(t, -\hat{p}_x)_\uparrow  + i \mathcal{O}_{N\pi}(t, \hat{p}_y)_\uparrow - i \mathcal{O}_{N\pi}(t,-\hat{p}_y)_\uparrow -\mathcal{O}_{N\pi}(t, \hat{p}_z)_\downarrow + \mathcal{O}_{N\pi}(t, -\hat{p}_z)_\downarrow,
\end{aligned}
\end{equation}
which are also the operators used in \cite{Barca_2023,Lang_2017,Prelovsek:2016iyo}. For the negative sector, we obtain 
\begin{equation}
\begin{aligned}
    \mathcal{O}^{(001)}_{N\pi, G_{1u}}(\boldsymbol 0, t)_{\uparrow} = \mathcal{O}_{N\pi}(t, \hat{p}_x)_{\uparrow} + \mathcal{O}_{N\pi}(t, -\hat{p}_x)_\uparrow  + \mathcal{O}_{N\pi}(t, \hat{p}_y)_\uparrow + \mathcal{O}_{N\pi}(t,-\hat{p}_y)_\uparrow +\mathcal{O}_{N\pi}(t, \hat{p}_z)_\uparrow - \mathcal{O}_{N\pi}(t, -\hat{p}_z)_\uparrow \\
    \mathcal{O}^{(001)}_{N\pi, G_{1u}}(\boldsymbol 0, t)_{\downarrow} = \mathcal{O}_{N\pi}(t, \hat{p}_x)_{\downarrow} + \mathcal{O}_{N\pi}(t, -\hat{p}_x)_\downarrow  + \mathcal{O}_{N\pi}(t, \hat{p}_y)_\downarrow + \mathcal{O}_{N\pi}(t,-\hat{p}_y)_\downarrow +\mathcal{O}_{N\pi}(t, \hat{p}_z)_\downarrow - \mathcal{O}_{N\pi}(t, -\hat{p}_z)_\downarrow,
\end{aligned}
\end{equation}
which is the same as described in \cite{Prelovsek:2016iyo}.
\section{Summary of the fit results}\label{app:fit_result}

All ensembles with at least 60 independent configurations (ensemble 4, D, 9) are fitted correlated in the positive parity channel. We need even larger statistics for the negative parity channel to obtain stable correlated fits. Therefore, we only fit ensemble 4 with correlated fits in this sector.  We will explore the thinning of correlation functions to reduce this problem in future work. \autoref{tab:results} shows the fit results for all ensembles and all effective mass curves. The table consists of two sub-tables, each consisting of 8 columns, starting with the fit tag, showing which effective mass curve is fitted, where N2N denotes the nucleon 2-point function, and the tags starting with GEVP are the individual modes of the GEVP. The next column displays the fit range of the chosen fit with the parameters shown in the following two columns. Here, we do not list the overlap factor $A$ results since we overestimate the error due to a non-symmetric distribution of the values for individual jackknife samples. The $\sigma_{\tau-1}$ shows the value from the extrapolation criterion introduced in \autoref{eq:extrapol_check}, which measures the tension between the first data point not included in the fit and the value we expect from the fitted model. In general, we aim for this value to be below $2$. However, there are some cases for which we found no fit range fulfilling this criterion, especially in the negative parity channel. The reduced $\chi^2$ value defined by the $\chi^2$ value over the degrees of freedom is shown in the next column. 
We choose the reduced $\chi^2$ over the p-value since we use a mixture of correlated and uncorrelated fits. The last column shows which fit is correlated (c) or uncorrelated (uc). 

\begin{table}[h!]
    \centering
    \footnotesize
    \parbox{.45\linewidth}{
    \begin{tabular}{|l|l|l|l|l|l|l|}
        \hline
        \multicolumn{7}{|c|}{\textbf{positive parity}}  \\  
        \hline
        Tag & range & $a E$ & $a E^{\text{ex}}$ & $\sigma_{\tau_0-1}$ & $\chi^2/\text{dof}$ & type \\
        \hline
        \hline
        \multicolumn{7}{|l|}{Ensemble-4} \\
        \hline
        GEVP0 & 4 - 12 & 0.6358(24) & 1.83(23) & 2.23 & 0.93 & c\\
        GEVP1 & 4 - 10 & 0.9458(55) & 1.86(21) & 1.75 & 1.73 & c\\
        GEVP2 & 4 - 10 & 0.9857(80) & 1.94(54) & 0.61 & 0.35 & c\\
        N2N & 5 - 12 & 0.6375(32) & 1.62(29) & 1.40 & 1.18 & c\\
        \hline
        \hline
        \multicolumn{7}{|l|}{Ensemble-D} \\
        \hline
        GEVP0 & 4 - 12 & 0.6244(23) & 1.48(16) & 0.87 & 0.25 & c\\
        GEVP1 & 4 - 12 & 0.9004(29) & 1.59(10) & 0.30 & 0.88 & c\\
        GEVP2 & 4 - 12 & 0.9362(35) & 1.60(10) & 0.43 & 0.52 & c\\
        N2N & 5 - 12 & 0.6248(24) & 1.44(15) & 0.93 & 0.24 & c\\
        \hline
        \hline
        \multicolumn{7}{|l|}{Ensemble-L} \\
        \hline
        GEVP0 & 6 - 18 & 0.4511(42) & 1.02(12) & 1.17 & 0.47 & uc\\
        GEVP1 & 6 - 18 & 0.6123(52) & 1.11(12) & 0.73 & 0.71 & uc\\
        GEVP2 & 6 - 18 & 0.6839(53) & 1.16(12) & 0.46 & 0.31 & uc\\
        N2N & 7 - 18 & 0.4510(42) & 1.02(12) & 1.15 & 0.50 & uc\\
        \hline
        \hline
        \multicolumn{7}{|l|}{Ensemble-9} \\
        \hline
        GEVP0 & 5 - 17 & 0.4622(26) & 0.898(71) & -1.11 & 1.26 & c\\
        GEVP1 & 6 - 15 & 0.661(18) & 0.877(90) & 0.50 & 0.69 & c\\
        GEVP2 & 5 - 15 & 0.7151(65) & 1.044(70) & -3.28 & 1.75 & c\\
        N2N & 6 - 17 & 0.4621(32) & 0.851(63) & -1.81 & 1.58 & c\\
        \hline
        \hline
        \multicolumn{7}{|l|}{Ensemble-1} \\
        \hline
        GEVP0 & 3 - 14 & 0.5909(41) & 1.51(23) & -0.31 & 0.19 & uc\\
        GEVP1 & 3 - 10 & 0.8260(58) & 1.61(18) & -0.87 & 0.15 & uc\\
        GEVP2 & 3 - 10 & 0.8432(70) & 1.51(13) & -2.00 & 0.16 & uc\\
        N2N & 4 - 15 & 0.5918(44) & 1.44(21) & -0.54 & 0.13 & uc\\
        \hline
        \hline
        \multicolumn{7}{|l|}{Ensemble-3} \\
        \hline
        GEVP0 & 2 - 10 & 0.5919(40) & 1.433(87) & 2.15 & 0.76 & uc\\
        GEVP1 & 4 - 10 & 0.811(31) & 1.05(30) & -1.87 & 0.29 & uc\\
        GEVP2 & 3 - 10 & 0.8521(68) & 1.53(14) & -2.00 & 0.12 & uc\\
        N2N & 3 - 11 & 0.5930(42) & 1.406(83) & 2.17 & 0.83 & uc\\
        \hline
        \hline
        \multicolumn{7}{|l|}{Ensemble-C} \\
        \hline
        GEVP0 & 4 - 18 & 0.5475(88) & 1.33(24) & 0.80 & 0.48 & uc\\
        GEVP1 & 4 - 18 & 0.6856(86) & 1.48(21) & 0.87 & 0.38 & uc\\
        GEVP2 & 4 - 18 & 0.7076(93) & 1.47(22) & 0.76 & 0.37 & uc\\
        N2N & 5 - 18 & 0.5473(90) & 1.31(24) & 0.73 & 0.45 & uc\\
        \hline
    \end{tabular}
    }
    \parbox{0.45\linewidth}{
    \begin{tabular}{|l|l|l|l|l|l|l|}
        \hline
        \multicolumn{7}{|c|}{\textbf{negative parity}}  \\   
        \hline
        Tag & range & $a E$ & $a E^{\text{ex}}$ & $\sigma_{\tau_0-1}$ & $\chi^2/\text{dof}$ & type \\
        \hline
        \hline
        \multicolumn{7}{|l|}{Ensemble-4} \\
        \hline
        GEVP0 & 4 - 9 & 0.7841(35) & 1.64(19) & 1.44 & 2.31 & c\\
        GEVP1 & 4 - 9 & 0.971(26) & 1.59(85) & 0.42 & 0.53 & c\\
        GEVP2 & 3 - 7 & 1.0058(82) & 3.19(13) & 2.96 & 0.44 & c\\
        N2N & 4 - 10 & 0.928(17) & 2.2(1.3) & -0.42 & 0.03 & c\\
        \hline
        \hline
        \multicolumn{7}{|l|}{Ensemble-D} \\
        \hline
        GEVP0 & 4 - 7 & 0.7782(27) & 1.59(17) & 0.57 & 0.00 & uc\\
        GEVP1 & 3 - 5 & 0.8958(38) & 1.523(51) & -8.47 & 0.00 & uc\\
        GEVP2 & 2 - 5 & 0.918(23) & 1.81(19) & 3.79 & 0.55 & uc\\
        N2N & 3 - 7 & 0.865(24) & 1.58(16) & 3.67 & 0.83 & uc\\
        \hline
        \hline
        \multicolumn{7}{|l|}{Ensemble-L} \\
        \hline
        GEVP0 & 4 - 8 & 0.5732(80) & 1.13(11) & 0.38 & 0.41 & uc\\
        GEVP1 & 4 - 7 & 0.6313(87) & 1.32(12) & 1.38 & 0.14 & uc\\
        GEVP2 & 2 - 6 & 0.690(52) & 1.19(16) & -1.35 & 0.03 & uc\\
        N2N & 4 - 7 & 0.679(87) & 1.18(40) & -0.93 & 0.04 & uc\\
        \hline
        \hline
        \multicolumn{7}{|l|}{Ensemble-9} \\
        \hline
        GEVP0 & 4 - 10 & 0.5698(33) & 0.983(44) & -0.41 & 0.06 & uc\\
        GEVP1 & 4 - 10 & 0.686(50) & 1.23(62) & 0.63 & 0.01 & uc\\
        GEVP2 & 4 - 10 & 0.7176(52) & 1.138(57) & -0.19 & 0.12 & uc\\
        N2N & 5 - 11 & 0.627(42) & 1.03(32) & 0.75 & 0.02 & uc\\
        \hline
        \hline
        \multicolumn{7}{|l|}{Ensemble-1} \\
        \hline
        GEVP0 & 4 - 10 & 0.7064(53) & 1.62(52) & 0.36 & 0.08 & uc\\
        GEVP1 & 3 - 8 & 0.8399(64) & 1.45(10) & -3.43 & 0.04 & uc\\
        GEVP2 & 3 - 6 & 0.8(1.4) & 1.1(1.9) & -1.17 & 0.06 & uc\\
        N2N & 4 - 10 & 0.802(98) & 1.39(77) & -0.36 & 0.18 & uc\\
        \hline
        \hline
        \multicolumn{7}{|l|}{Ensemble-3} \\
        \hline
        GEVP0 & 4 - 8 & 0.703(11) & 1.13(28) & -1.05 & 0.25 & uc\\
        GEVP1 & 2 - 6 & 0.837(49) & 1.53(36) & 0.20 & 0.26 & uc\\
        GEVP2 & 3 - 6 & 0.8661(55) & 3.0(1.1) & 0.77 & 0.19 & uc\\
        N2N & 4 - 7 & 0.800(29) & 1.85(61) & 0.39 & 0.15 & uc\\
        \hline
        \hline
        \multicolumn{7}{|l|}{Ensemble-C} \\
        \hline
        GEVP0 & 3 - 7 & 0.621(11) & 1.24(11) & -1.66 & 0.48 & uc\\
        GEVP1 & 2 - 6 & 0.6825(55) & 1.412(40) & -1.86 & 1.06 & uc\\
        GEVP2 & 2 - 5 & 0.723(93) & 1.35(31) & 1.46 & 0.33 & uc\\
        N2N & 3 - 6 & 0.711(82) & 1.35(28) & 1.52 & 0.32 & uc\\
        \hline
    \end{tabular}
    }
    \caption{Summary over all excited state fits of the GEVP and two-point effective mass curves with the fit form from \autoref{eq:fit_form} for the different ensembles. The last column denotes whether a fit is correlated (c) or uncorrelated (uc).}
    \label{tab:results}
\end{table}

\begin{figure}[h!]
    \centering
    \includegraphics[width=\textwidth]{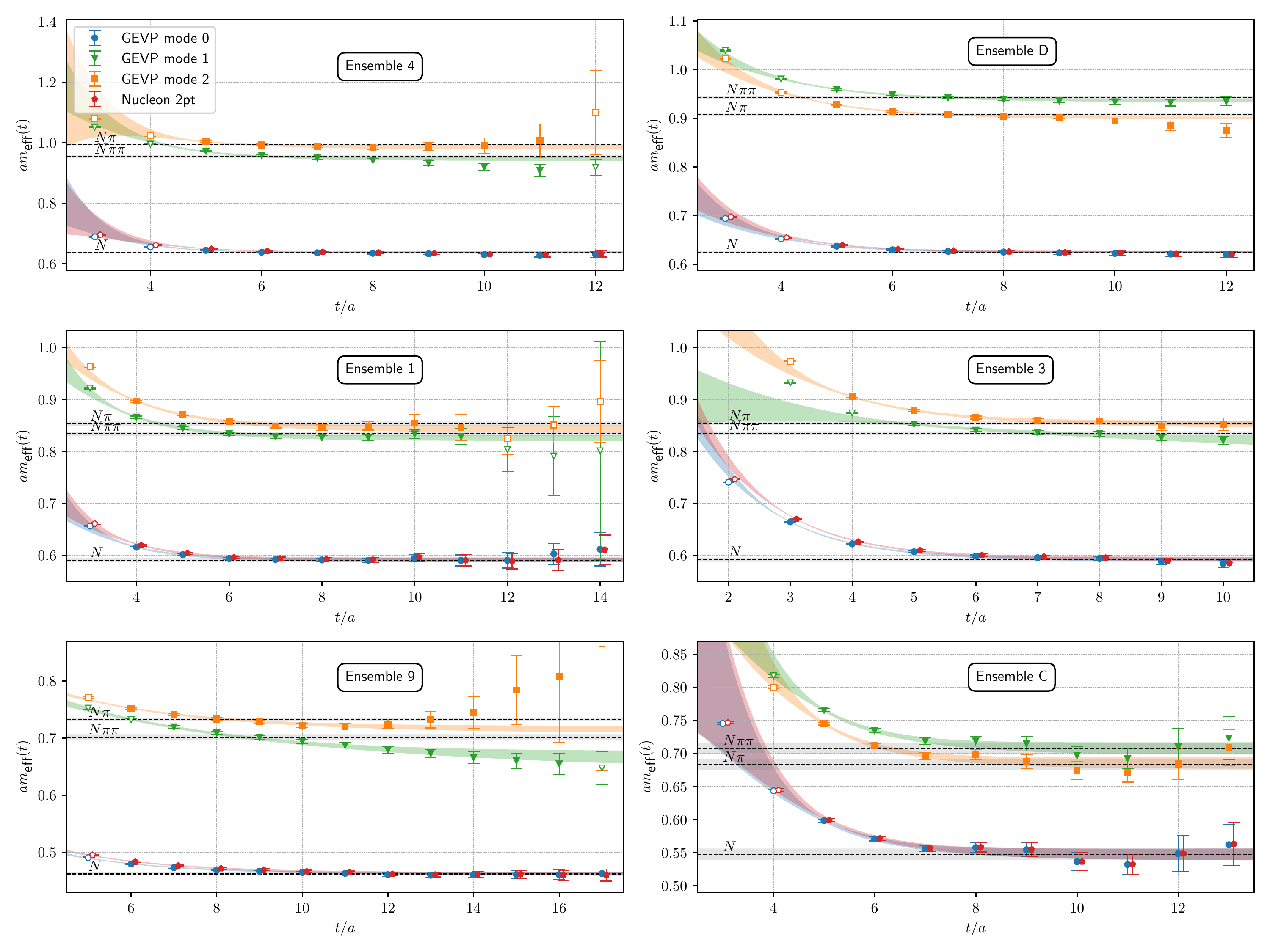}
    \caption{Overview plots of the remaining ensembles showing the same results as depicted in \autoref{fig:gevp-L} for all other ensembles of this study.}
    \label{fig:all_gevp}
\end{figure}

\begin{figure}[h!]
    \centering
    \includegraphics[width=\textwidth]{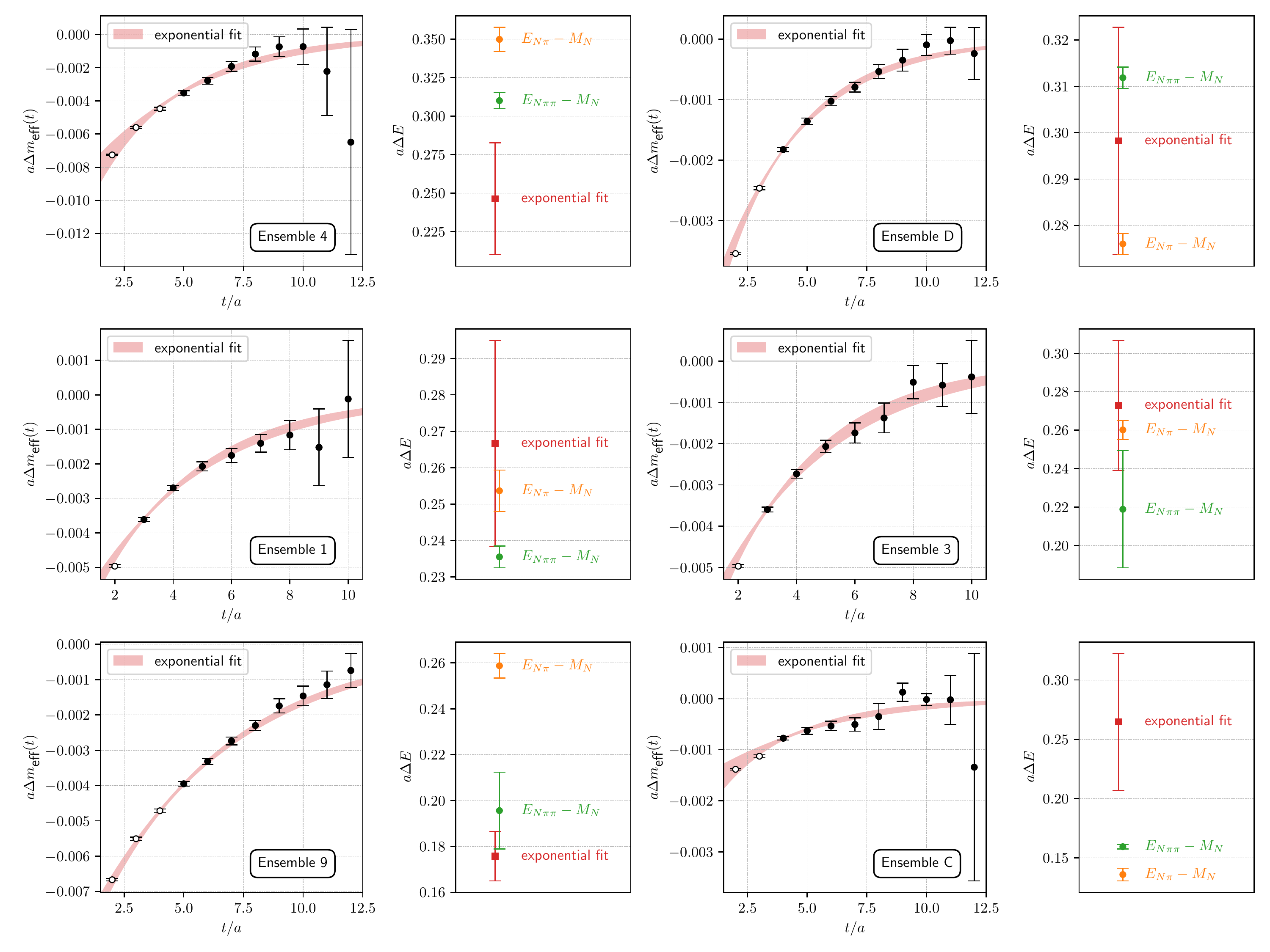}
    \caption{Overview plots of the remaining ensembles showing the same results as depicted in \autoref{fig:diff-L} for all other ensembles of this study.}
    \label{fig:all_diff}
\end{figure}

\begin{figure}[h!]
    \centering
    \includegraphics[width=\textwidth]{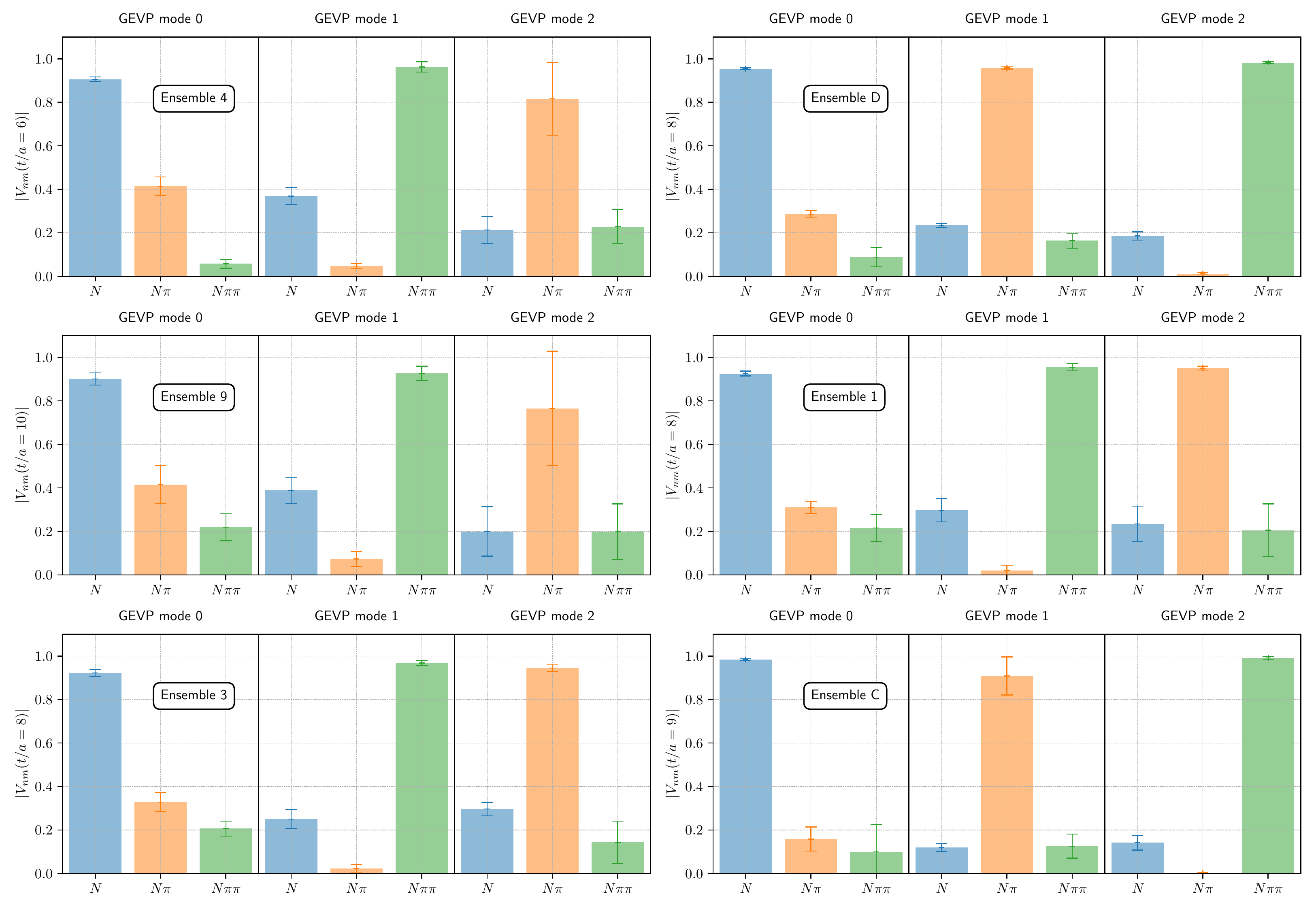}
    \caption{Overview plot of all GEVP eigenvectors. The plots depict the same eigenvectors as \autoref{fig:evec-L} for the remaining ensembles of this study. The reference time slices can vary between the different ensembles and are chosen so that the effective mass is converged to the state's mass while still having a moderate signal-to-noise ratio.}
    \label{fig:all_evec}
\end{figure}

\begin{figure}[h!]
    \centering
    \includegraphics[width=\linewidth]{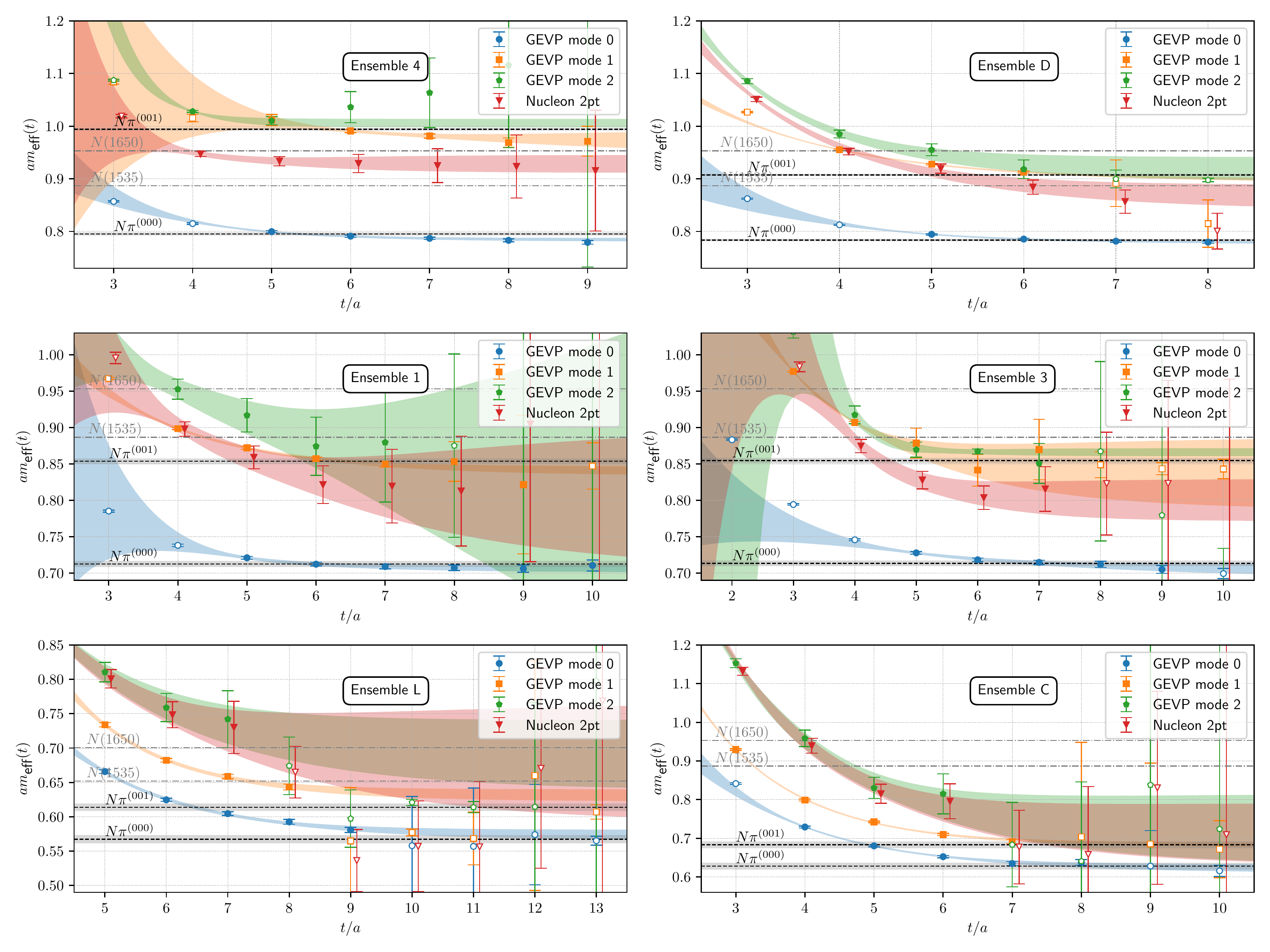}
    \caption{Overview over the GEVP effective masses of the remaining ensembles in the negative parity sector. The plots are similar to \autoref{fig:ngevp-9}}
    \label{fig:ngevp-all}
\end{figure}

\begin{figure}[h!]
    \centering
    \includegraphics[width=\linewidth]{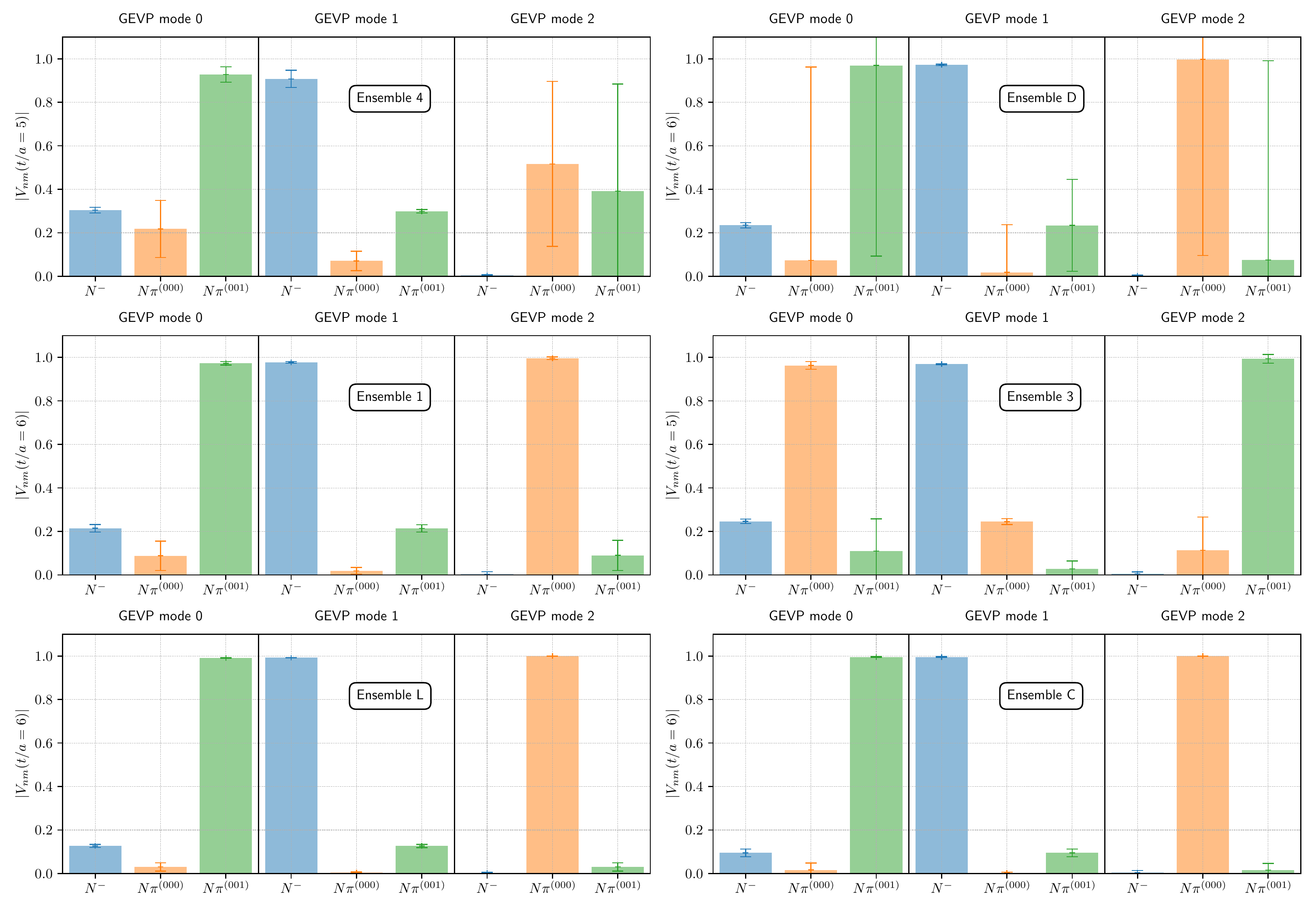}
    \caption{Overview plot of all negative parity GEVP eigenvectors. The plots depict the same eigenvectors as \autoref{fig:ngevp-9-evec} for the remaining ensembles of this study. The reference time slices can vary between the different ensembles and are chosen such that the effective mass is converged to the state's mass while still having a moderate signal-to-noise ratio.}
    \label{fig:ngevp-all-evec}
\end{figure}

\begin{figure}
    \centering
    \includegraphics[width=\linewidth]{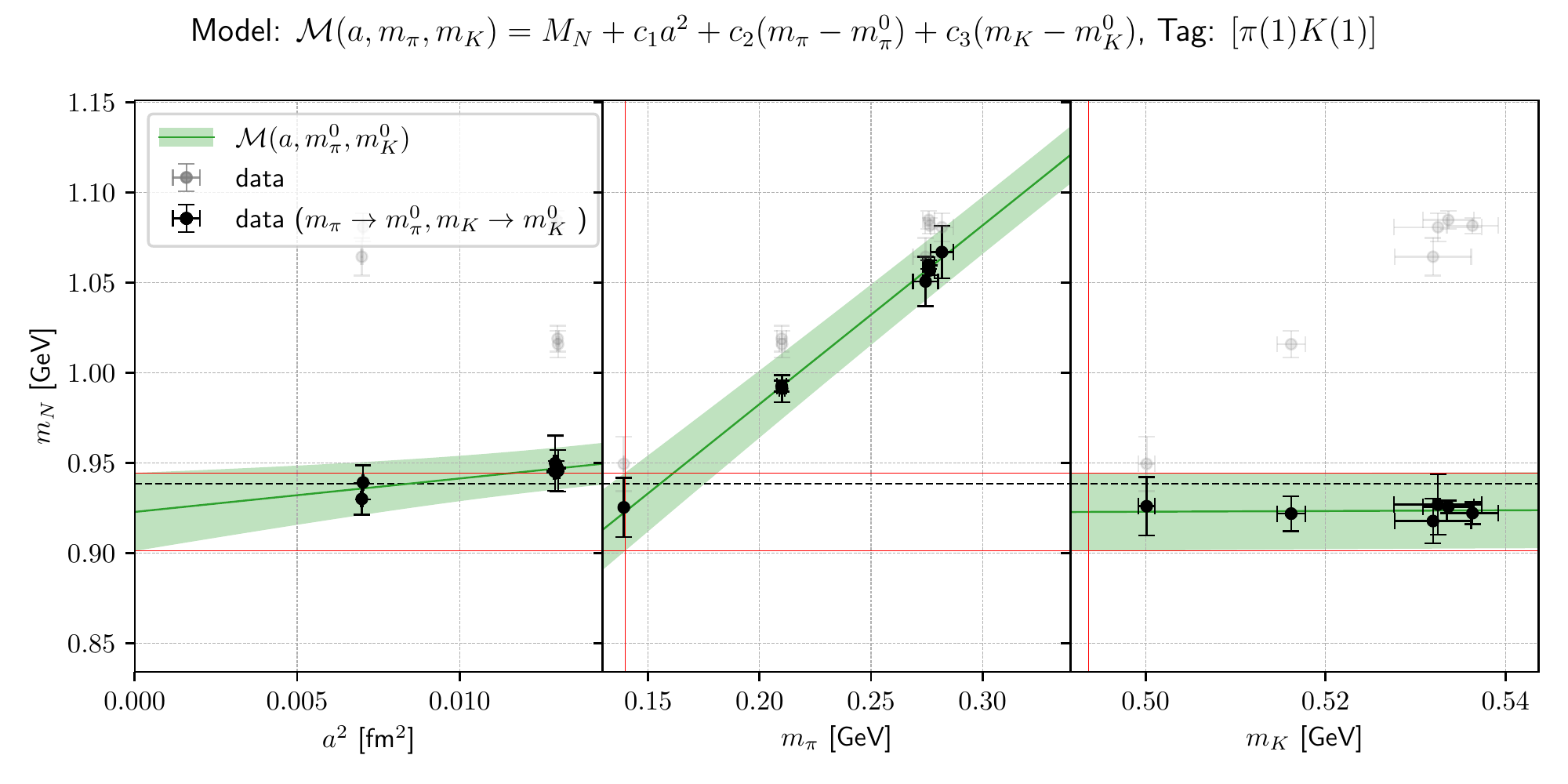}
    \includegraphics[width=\linewidth]{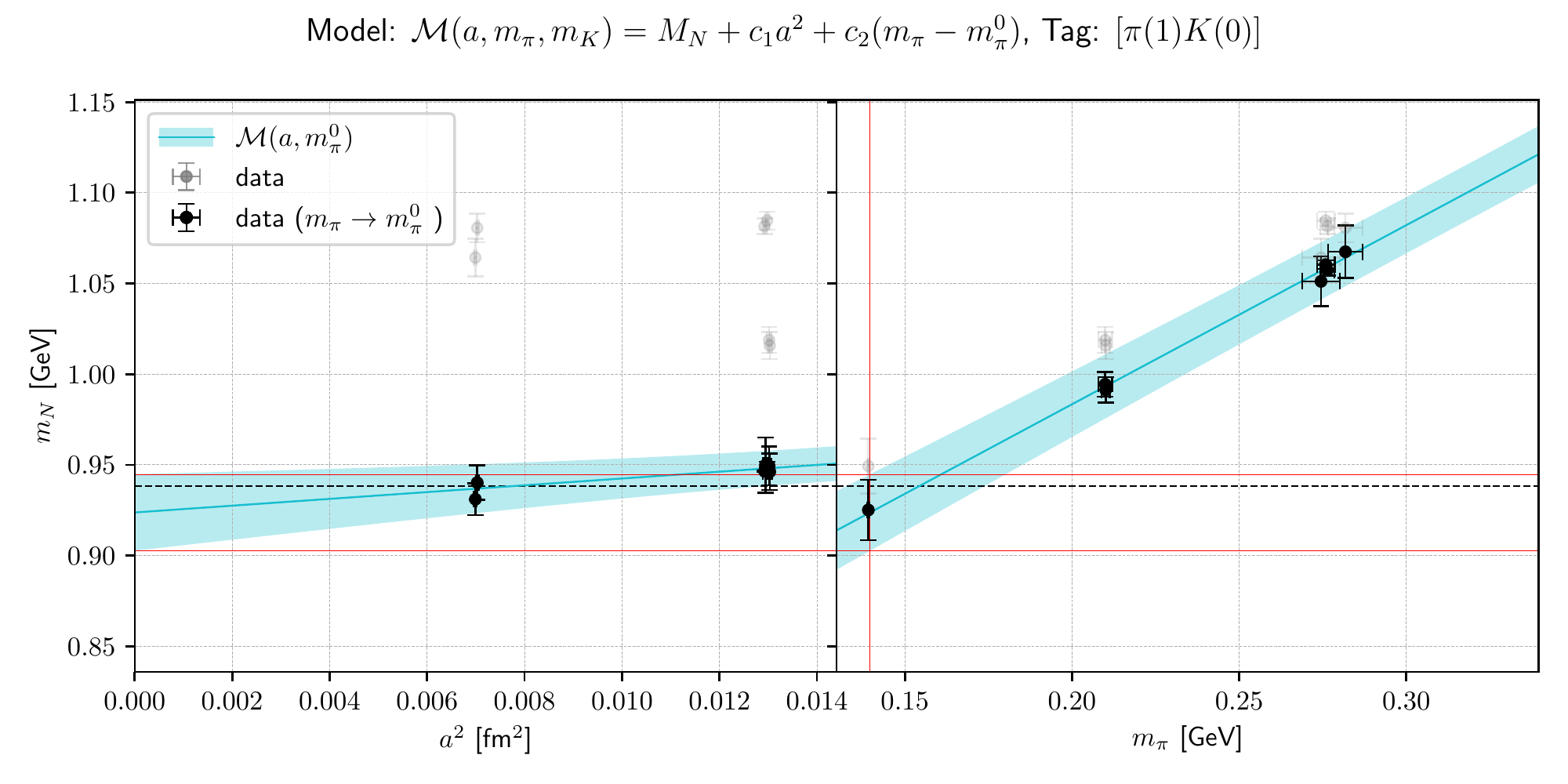}
    \caption{Overview of the linear models for the continuum and physical point extrapolation. The plots are similar to \autoref{fig:limit_example}.}
    \label{fig:lin_model_summary}
\end{figure}

\begin{figure}
    \includegraphics[width=0.8\linewidth]{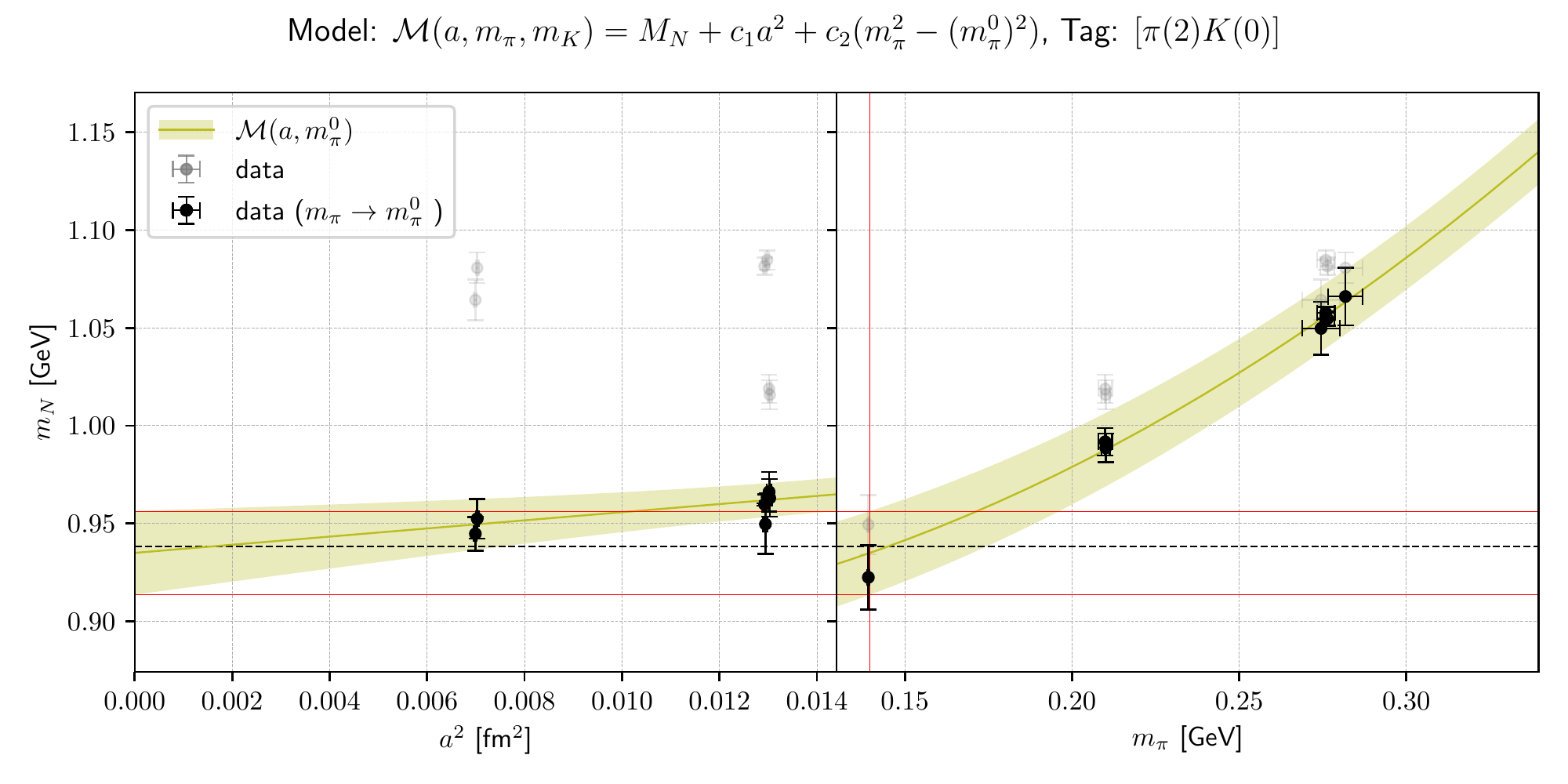}
    \includegraphics[width=0.8\linewidth]{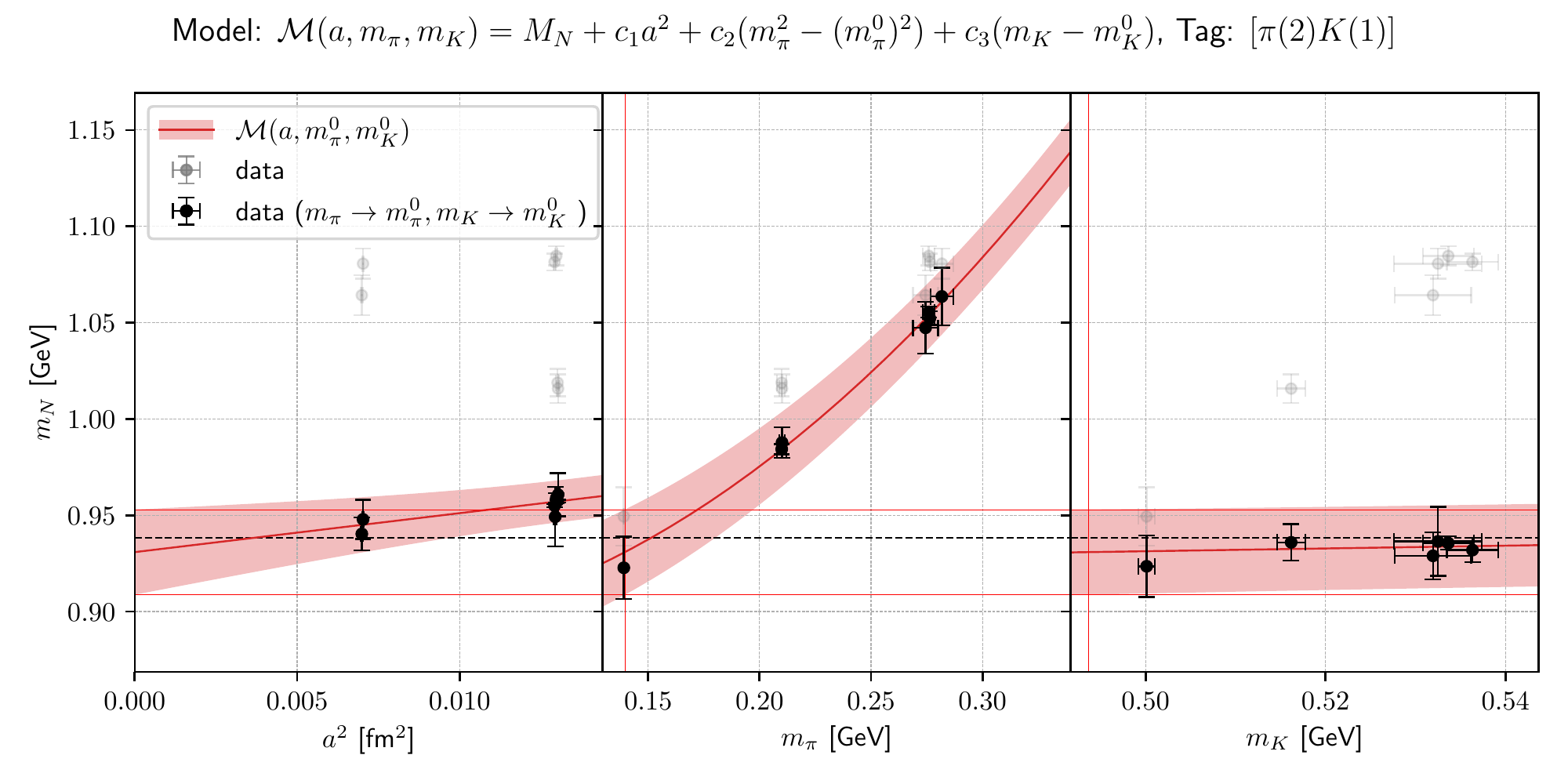}
    \includegraphics[width=0.8\linewidth]{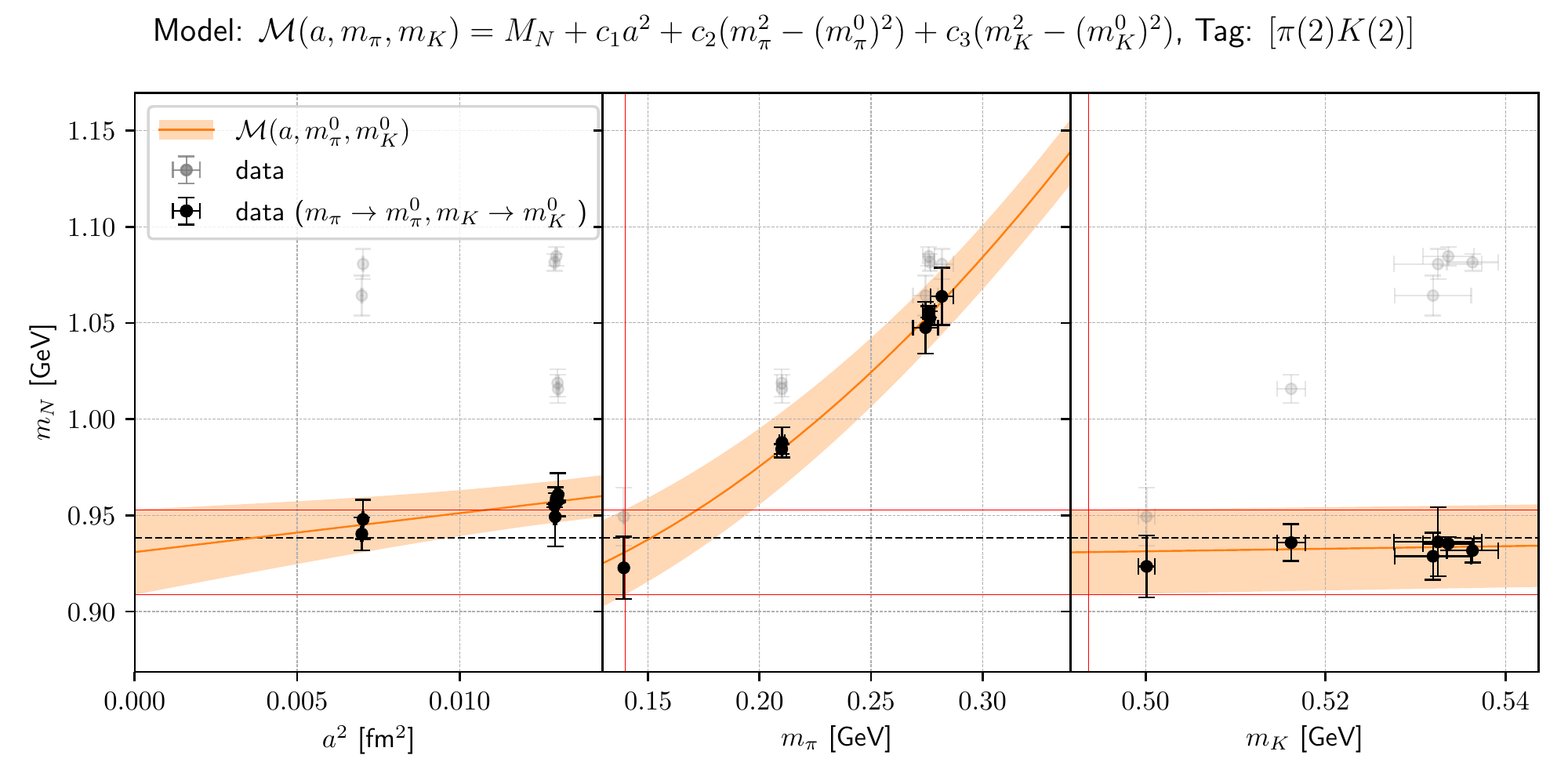}
    \caption{Overview of the quadratic models for the continuum and physical point extrapolation. The plots are similar to \autoref{fig:limit_example}.}
    \label{fig:quad_model_summary}
\end{figure}

\begin{figure}
    \includegraphics[width=0.8\linewidth]{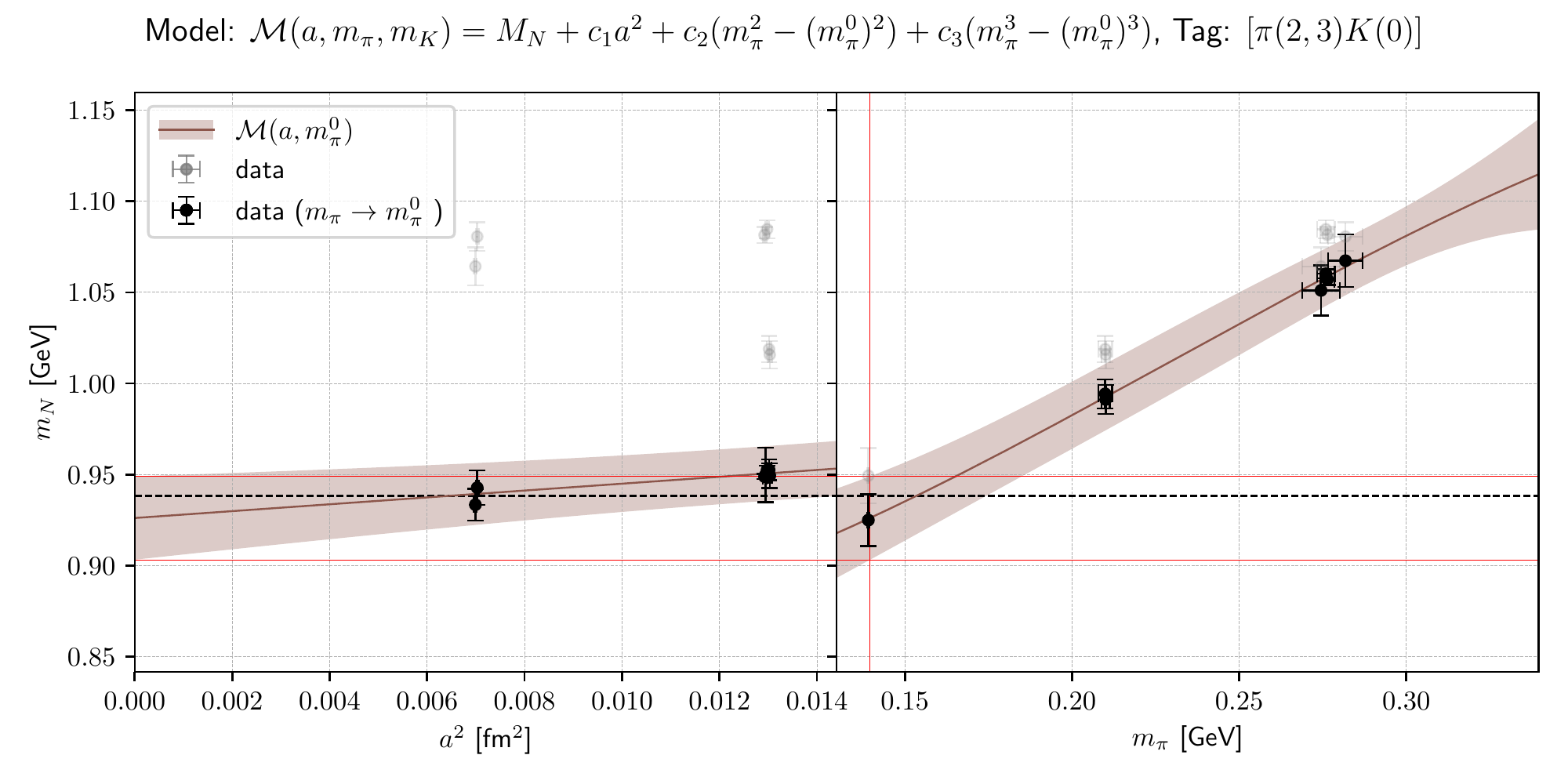}
    \includegraphics[width=0.8\linewidth]{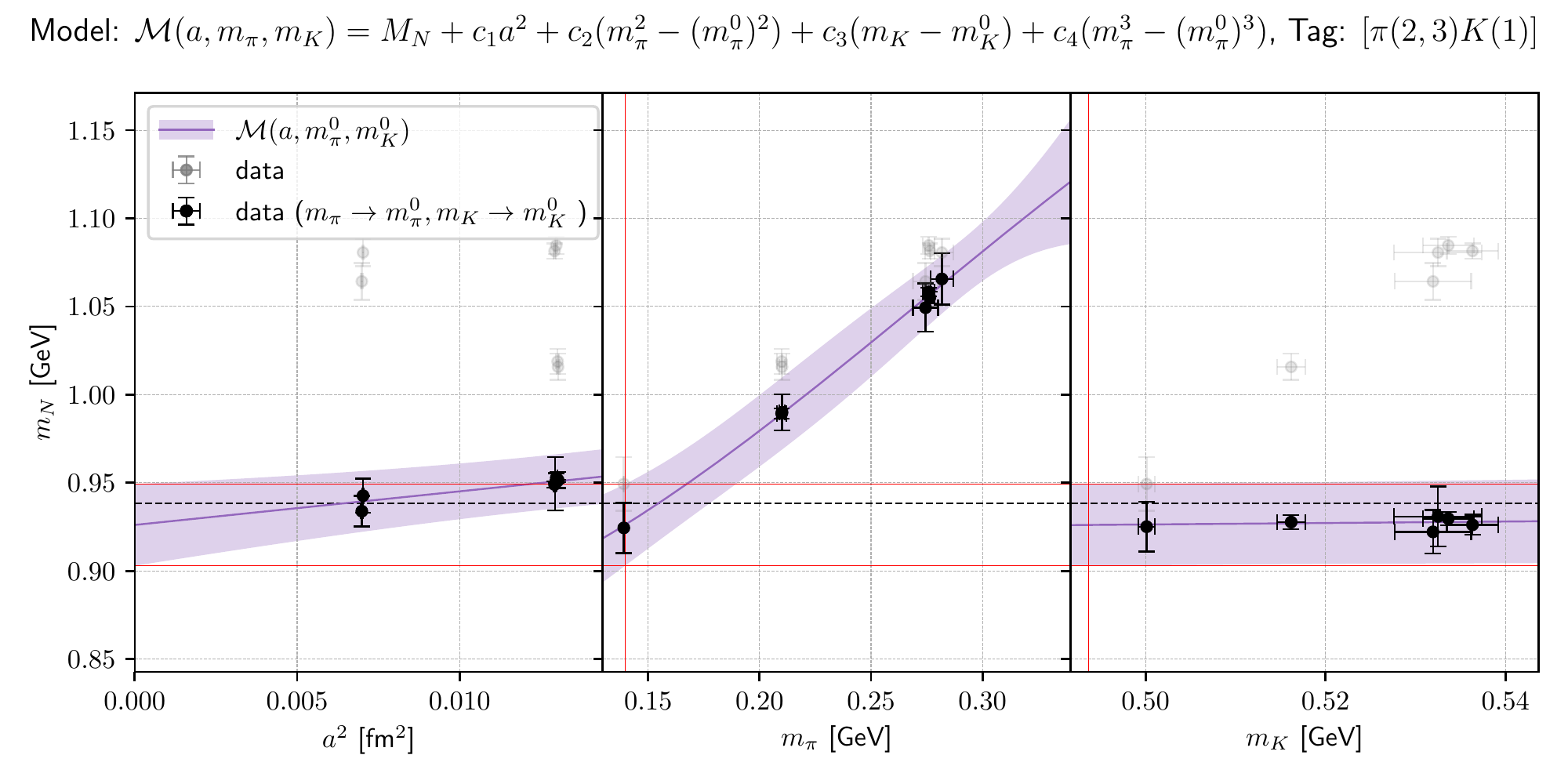}
    \includegraphics[width=0.8\linewidth]{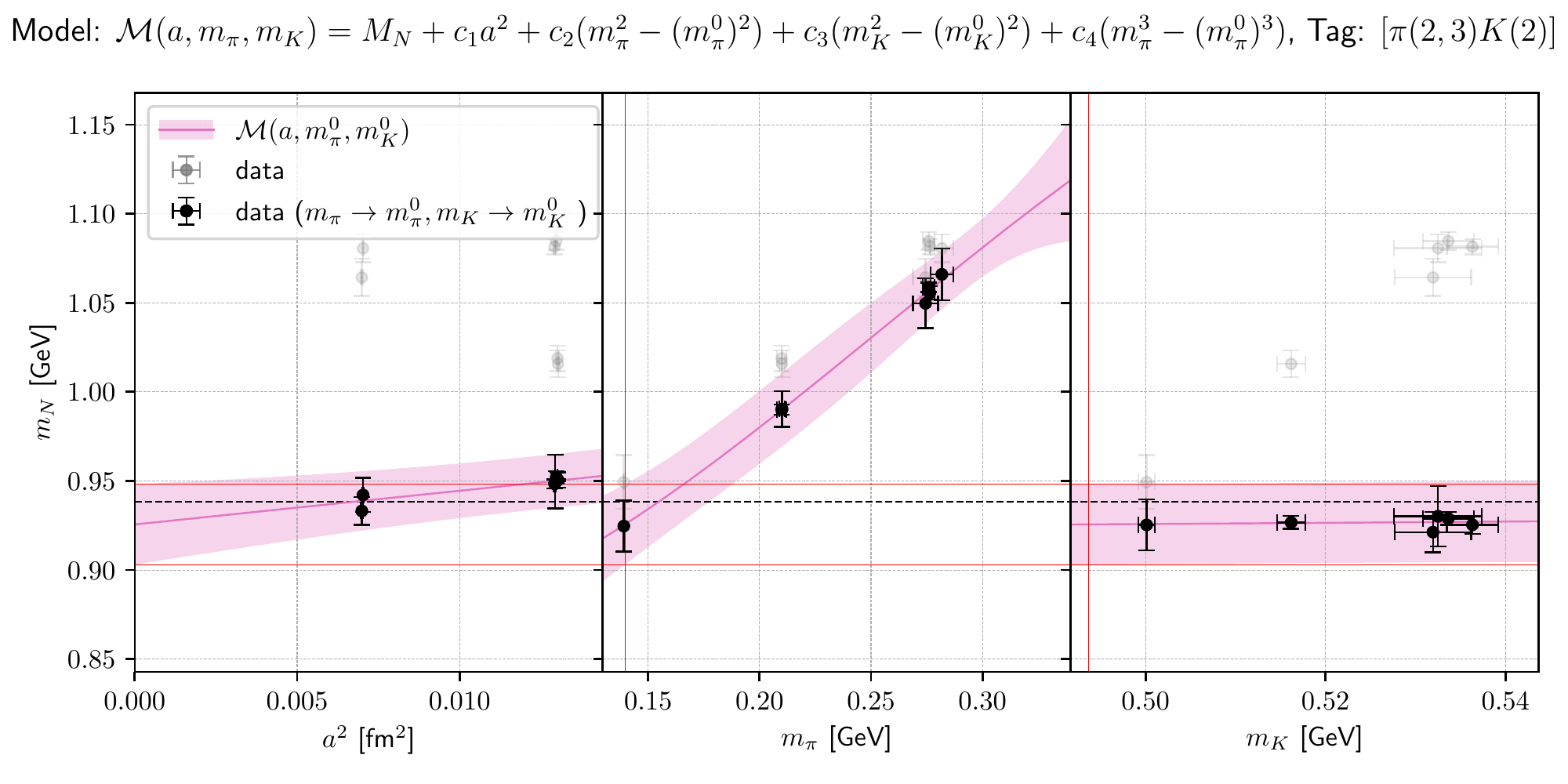}
    \caption{Overview of the quadratic models with a cubic term for the continuum and physical point extrapolation. The plots are similar to \autoref{fig:limit_example}.}
    \label{fig:quad_cub_model_summary}
\end{figure}

\end{document}